\documentclass[aps,prd,11pt,reprint,nofootinbib,showkeys,superscriptaddress]{revtex4-2} 
\usepackage{amsmath,amssymb,amsthm,mathrsfs}
\usepackage{latexsym,amscd,amsbsy,amsfonts}
\usepackage{graphicx}
\usepackage{ragged2e}
\usepackage{caption}
\usepackage[utf8]{inputenc}
\usepackage{marvosym}
\usepackage{float}
\usepackage{slashed}
\usepackage{cancel}
\usepackage{multirow}
\usepackage{soul}
\usepackage{physics}
\usepackage{comment}
\usepackage[usenames,dvipsnames]{xcolor}
\usepackage[
      colorlinks=true,
      linktocpage=true,
      breaklinks=true,
      linkcolor=Aquamarine,
      urlcolor=Purple,
      filecolor=black,
      citecolor=Purple,
      menucolor=black,
      pdfstartview=FitV,
      bookmarksopen=true
      ]{hyperref}
\usepackage{tikz}
\usepackage{dsfont}
\usepackage{ragged2e}

\usepackage[usenames,dvipsnames]{xcolor}
\definecolor{emerald}{rgb}{0.0, 0.0, 0.0}
\newcommand\jao[1]{{\color{emerald} #1}}

\begin{document}

\title{Quantum entanglement of Hawking-Partner modes in expanding cavities}

\author{José Manuel Montes-Armenteros}
\email{jmmontes@correo.ugr.es}
\affiliation{ Universidad de Granada, Granada-18071, Spain}

\author{Javier Olmedo}
\email{javolmedo@ugr.es}
\affiliation{Departamento de F\'isica Te\'orica y del Cosmos, Universidad de Granada, Granada-18071, Spain}

\begin{abstract}
This article investigates quantum entanglement generated within a one-dimensional cavity where one boundary undergoes prescribed acceleration, a setup designed to mimic aspects of Hawking radiation. We quantify quantum correlations using logarithmic negativity for bipartitions where subsystem $A$ is a given mode and subsystem $B$ is the rest of the system. For initial pure states, we also consider a given mode and reconstruct its partner using the Hotta-Sch\"utzhold-Unruh formula, obtaining identical results. \jao{Interestingly, this last method offers notable computational efficiency. However, partner modes do not commute, due to the nontrivial multimode entanglement structure. Hence, a pairwise description will not be suitable for describing the full system. Besides, our findings reveal that the expanding cavity effectively acts as a squeezing device, with Hawking-partner pairs largely behaving as two-mode squeezed states. We checked that, in our setting, purification of Hawking modes is predominantly a low-energy process, with high-energetic particles contributing negligibly to the partner modes. Indeed, in both small and large acceleration regimes of the boundaries, quantum entanglement decreases toward the ultraviolet modes, indicating that higher-energy particles are more challenging to entangle and hence less probable to contribute in the purification process. } Besides the initial vacuum state, we also consider one-mode squeezed and two-mode squeezed states, in order to confirm if quantum entanglement can be stimulated. Moreover, we analyze its robustness against initial thermal noise. Our analysis is based on numerical simulations and does not assume any approximation beyond the validity of our numerical algorithms.  \jao{We conclude with a discussion about the possible implementation and observation of our results in the laboratory.}

\end{abstract}

\maketitle

\section{Introduction}

The phenomenon of Hawking radiation, originally predicted in the context of black hole physics, has profound implications for our understanding of quantum fields in curved spacetimes and the interplay between general relativity and quantum mechanics. Hawking’s seminal work demonstrated that black holes emit thermal radiation, leading to gradual loss of mass and, ultimately, evaporating, perhaps leaving a final remnant. Despite its fundamental importance, direct experimental observation of Hawking radiation from astrophysical black holes remains elusive because of its extremely low intensity and temperature for typical black hole masses. Besides, the emitted radiation depends only on the black hole mass, charge, and angular momentum, meaning that any detailed information about the matter that formed the black hole would be lost, leaving behind a mixed state of radiation. This outcome contradicts the fundamental principle of unitarity in quantum mechanics, namely, that information must be preserved under time evolution. Instead, if a pure quantum state evolves into a mixed state, unitarity will not be satisfied. This problem has spurred decades of \jao{research} \cite{Unruh:2017uaw,Marolf:2017jkr,Modak:2014qja,Almheiri:2020cfm,Ashtekar:2025ptw}, with the precise mechanism by which this occurs remaining a subject of debate within the community. 

To circumvent these conceptual challenges, researchers have developed analogue systems that simulate the essential features of Hawking radiation in controlled laboratory settings. One of the first analogue systems suggested by Fulling and Davies considered an accelerating mirror \cite{Davies1977}. They showed that particular trajectories of the mirror are able to excite the field vacuum state into a thermal one. Later on it was shown that there exist a plethora of trajectories able to produce thermal particles \cite{Good:2013lca,Akal:2022qei}. Other examples of promising analogue systems where spontaneous Hawking radiation is expected are those based on Bose-Einstein condensates \cite{Garay1999,Garay2000,Barcelo2001,Steinhauer2015,Nova:2019,Kolobov2019,Ribeiro:2022gln}, quantum fluids of light~\cite{Carusotto2004,Nguyen2015}, electromagnetic waveguides~\cite{Schutzhold:2004tv}, and optical fibers~\cite{Philbin:2007ji,Belgiorno2010,Rubino2011,Drori2018,Agullo2021,Brady:2022ffk}. See also Ref.~\cite{Barcelo2005} for a detailed review. 

Moving mirrors provide an excellent theoretical setting for the study of Hawking radiation and the information loss problem. There have been attempts to explain how information is
restored via entanglement of the early-time Hawking radiation with vacuum fluctuations in the far future, when the mirror returns to inertial motion~\cite{hotta2015Partner}. However, as stated in Ref.~\cite{Wald_2019}, this scenario also suffers from similar limitations as the usual last-burst models, where Hawking radiation is purified by a final, very high energetic burst: there appears to be an indirect energy cost for purification. See also \cite{osawa2024finalburstmovingmirror} for a more recent discussion of this issue. It seems that the heart of all these problems lies in the requirement that information loss must be solved within the weak-field approximation, where semiclassical physics remains valid. Actually, in the context of more realistic, evaporating black holes, it was recently shown in Ref.~\cite{Agullo_2024} that partner modes explore spacetime regions where general relativity ceases to be valid, making unprovable that semiclassical physics is enough to explain the information loss problem. 

A very close approach recently considered within this context is a quantum field confined in a one-dimensional cavity, where the motion of one or two boundaries can mimic the effect of a gravitational horizon \cite{GarciaMartin-Caro:2023jjq,GarciaMartin-Caro:2024qpk,GarciaMartin-Caro:2025xgr}. Here, the acceleration of a boundary induces a change in the vacuum state of the field, leading to particle creation and the emergence of a nearly thermal spectrum, closely resembling the Hawking effect. One of the advantages of this setting is their experimental feasibility using coplanar wave guides ended in superconducting quantum interference devices (SQUIDs)~\cite{Wilson_2011,Johansson:2009zz,Johansson:2010vqd,Doukas:2014bja,Pasi2013}. In fact, they have been considered in other configurations in order to simulate black hole-like geometries \cite{Terrones:2021bjp,Maceda:2025jti}. 

In this work, following that framework, we investigate quantum correlations by computing the logarithmic negativity ---a well-established measure of entanglement--- using both general expressions for a system of $N$ interacting modes, taking a bipartition with a given mode as subsystem $A$ and the rest of the modes as subsystem $B$, and alternatively by tracing out all modes except one, corresponding to a Hawking mode, and explicitly constructing its partner, inspired by Hotta-Sch\"utzhold-Unruh in Ref.~\cite{hotta2015Partner} (see also Refs. \cite{Trevison:2018ear,Tomitsuka:2019ehm} for a more recent treatment including its application to a field theory in a box). \jao{We checked that this formula offers considerable numerical efficiency. However, we will see that the nontrivial correlations between Hawking modes implies the noncommutativity of their partners, disabling the possibility of a pairwise description of the system in terms solely of partner-Hawking modes.} In addition, we study initial states different from the vacuum to understand if it is possible to stimulate quantum entanglement between pairs of particles, and also discuss the robustness of quantum entanglement against initial thermal noise, following the ideas of Refs. \cite{Agullo2021,Brady:2022ffk}, which are also based on Gaussian state theory \cite{weedbrook2012gaussian,Adesso2014GaussianStatesAndBeyond}. This approach and the operational advantages of the methods we adopt here allow us to quantify the entanglement structure of the resulting state and to clarify the role and properties of the Hawking partners that purify the state upon tracing out all other degrees of freedom. 

Our manuscript is organized as follows. In Sec. \ref{sec:qft} we provide the basic mathematical and physical details of a quantum field in a box with moving boundaries. We then discuss several properties of Gaussian states and quantum entanglement in Sec. \ref{sec:gauss-entangl}. In Sec. \ref{sec:partner} we apply the partner formula to purify a given mode after tracing out all other modes of the field. \jao{We carry out a detailed numerical study of the system in Sec. \ref{sec:numerics}, for small and large accelerations of the cavity. In both cases, we compute the logarithmic negativity between Hawking-partner modes and also for Hawking modes with the rest of the system, in the latter case including initial squeezed and thermal states. }We conclude in Sec. \ref{sec:concl}. We also added two Appendices.

\section{Classical and quantum aspects of a scalar field in a box with moving boundaries}\label{sec:qft}

Let us consider a massless, Klein-Gordon scalar field $\phi(t,x)$ inside a one-dimensional cavity, which satisfies the Dirichlet boundary conditions $\phi(t,x=f(t))=\phi(t,x=g(t))=0$. Here, $f(t)$ and $g(t)$ are the trajectories of the left and right boundaries, respectively. Then, there is a natural Fourier mode decomposition for the field given by
\begin{equation}\label{eq:fourier}
 \phi(t,x)=  \sum_{n=1}^{\infty} \phi_n(t)\sin \left[\frac{n \pi}{L(t)} \Big(x-f(t)\Big)\right],
\end{equation}
where $L(t)=g(t)-f(t)$ is the length of the cavity. 
The Fourier coefficients $\phi_n(t)$ satisfy the equations of motion
\begin{equation}
    \ddot{u}_n+ \sum_{m=1}^\infty R_{nm}\dot{u}_m +\sum_{m=1}^{\infty} S_{nm}u_m=0,
    \label{shortEoM}
\end{equation}
with
\begin{equation}\label{eq:RnmSnm-b}
    \begin{split}
          &S_{mn}=\delta_{mn}\bigg[\left(\frac{n\pi}{L}\right)^2\left(1-\dot f^2-\dot f\dot L-\frac{\dot L^2}{3}\right)\\
          & +\frac{\dot L^2}{2L^2}\left(1-\frac{2}{n\pi}\right)+\frac{\ddot L}{2 n\pi L}\bigg]+ ((-1)^{m+n}-1)\\
          & \times\bigg[\frac{2[\ddot fL+\ddot LL-2\dot f\dot L-2\dot L^2]m}{(m^2-n^2)\pi L^2}
          -\frac{8[\dot f\dot L+\dot L^2]mn^3}{(m^2-n^2)^2L^2}\bigg], \\
          &R_{mn}= -\delta_{mn}\frac{\dot L }{L}-(1-\delta_{mn})\frac{4[(-1)^{m+n}(\dot f+\dot L)-\dot f]mn}{(m^2-n^2)L}.
    \end{split}
\end{equation}
It is obvious that if $\dot L(t)\neq 0$ and/or $\dot f(t)\neq 0$, there will be dynamical mode mixing. We will see that this translates into correlations in the quantum theory and eventually to quantum entanglement. 

In what follows, we will complexify the space of solutions, but keeping in mind that (complex) Fourier modes must satisfy the reality conditions $\bar\phi_n(t) = \phi_n(t)$. 

With this in mind, we can compute complex solutions to the equations of motion above given some suitable initial data. Moreover, any solution will be a vector in the complex space of solutions with an infinite number of components (a pair configuration and velocity for each mode). Let us denote a solution ${\bf U}(t)$ as
\begin{eqnarray}\nonumber
    &&{\bf U}(t) = \Big(U_{1,\epsilon_1}(t),U_{2,\epsilon_2}(t), \ldots\Big),
\end{eqnarray}
such that $U_{n,\epsilon_n}(t)=\Big(\phi_n(t),\pi_n(t)\Big)$ and with $U_{n,\epsilon_n=0}(t)=\phi_n(t)$ and $U_{n,\epsilon_n=1}(t)=\pi_n(t)$, and where
\begin{align}\nonumber
& \pi_{n}={L}\dot \phi_{n}+\frac{\dot L}{2} \phi_{n} -2 \sum_{m}\left(1-\delta_{m n}\right)\frac{m n}{m^{2}-n^{2}} \\ 
& \times\left[{\dot f}\left((-1)^{m+n}-1\right)+{\dot L}(-1)^{m+n}\right]  \phi_{m}
\end{align}
are the conjugate momenta of $\phi_n$ with Poison algebra
\begin{equation}\label{eq:fourier-poisson}
\left\{\phi_{n}, \pi_{m} \right\}=2 \delta_{n m}.
\end{equation}
Moreover, let us consider any $\boldsymbol{U} (t) \in \cal{S}^{\mathbb{C}}$ with $\cal{S}^{\mathbb{C}}$ the complexified space of solutions. One can see that it is endowed with a natural Klein-Gordon product which is preserved under the evolution, i.e., a map $\cal{S}^{\mathbb{C}} \times \cal{S}^{\mathbb{C}} \rightarrow \mathbb{C}$. Given two complex solutions ${\bf U}^{(1)}(t)$ and ${\bf U}^{(2)}(t)$, this product is expressed as
\begin{align}\label{eq:KG-prod}
&\langle {\bf U}^{(1)}(t), {\bf U}^{(2)}(t)\rangle=\frac{i}{2} \sum_{n=1}^{\infty} \Big(\bar{\phi}_{n}^{(1)}(t) \pi_{n}^{(2)}(t)-\bar{\pi}_{n}^{(1)}(t) \phi_{n}^{(2)}(t)\Big).
\end{align}
This Klein-Gordon product is time independent on solutions, namely, it is unaffected by the choice of $t$. Besides, this product is not positive definite. Hence it cannot be used straightforwardly to endow our complexified space of solutions with an actual inner product. One must choose a subspace of the complex space of solutions where the Klein-Gordon product~\eqref{eq:KG-prod} is positive definite. This subspace ${\cal S}^+ \subset {\cal S}^{\mathbb{C}}$ is usually called the positive frequency sector of the theory. The complementary sector (where the inner product is negative) corresponds to the complex conjugate solutions of the positive frequency ones. 

From a practical perspective, we solve this set of equations \eqref{shortEoM} numerically, in the nonperturbative regime of the boundary configurations. We will adopt an explicit embedded Prince-Dormand-Runge-Kutta (8,9) method. Concrete details can be found in \cite{jolmedo2025}. In order to do so, we truncate the maximum number of modes $N$ and eventually extrapolate to the case $N\to\infty$. Concretely, we choose $N=256,\,512,\, 1024$, and adopt a Richardson extrapolation to obtain the limit $N\to\infty$ when necessary. We use this method to numerically evaluate in the asymptotic future (once the boundaries become stationary) the $in$-basis of complex, positive frequency solutions ${}^{in}{\bf u}^{(I)}(t)=\big({}^{in}u_{1,\epsilon_1}^{(I)}(t),{}^{in}u_{2,\epsilon_2}^{(I)}(t),\ldots\big)$ with  $I=1,2,\ldots$, and their complex conjugate. This basis is well adapted to the natural $in$ vacuum state at early times (asymptotic past), where the boundaries also remain stationary. In general, any basis $\left({\bf u}^{(I)}, \bar{{\bf u}}^{(I)}\right)$, with $I=1,2,\ldots$ of solutions is normalized with respect to the Klein-Gordon product \eqref{eq:KG-prod} as
\begin{equation}\label{eq:basis}
    \langle {\bf u}^{(I)}, {\bf u}^{\left(J\right)}\rangle=\delta^{I J}, \;\langle {\bf u}^{(I)}, \bar{{\bf u}}^{\left(J\right)}\rangle=0, \;\langle\bar{{\bf u}}^{(I)}, \bar{{\bf u}}^{\left(J\right)}\rangle=-\delta^{I J}.
\end{equation}

Hence, any real solution to the equations of motion can be expressed as 
\begin{equation}\label{eq:sol}
{\bf U}(t)=\sum_{I=1}^\infty a_I {\bf u}^{(I)}(t)+\bar a_I \bar{{\bf u}}^{(I)}(t), 
\end{equation}
where $\bar a_I$ and $a_I$ are the creation and annihilation variables defined as
\begin{equation}
a_{I}=\left\langle {\bf u}^{(I)}(t), {\bf U}(t)\right\rangle, \quad \bar{a}_{I}=-\left\langle\bar{{\bf u}}^{(I)}(t), {\bf U}(t)\right\rangle.
\end{equation}
They satisfy 
\begin{equation}\label{eq:amu-poisson}
\left\{a_{I}, \bar{a}_{J}\right\}=-i\delta^{IJ},\quad \left\{a_{I}, {a}_{J}\right\}=0=\left\{\bar a_{I}, \bar{a}_{J}\right\},
\end{equation}
provided the Fourier modes fulfill the Poisson algebra~\eqref{eq:fourier-poisson} and the basis of solutions normalized as in Eq.~\eqref{eq:basis}.\footnote{Besides, if we impose the Poisson algebra~\eqref{eq:amu-poisson}, we also obtain the closure conditions
\begin{equation}\label{eq:closure-cond}
\frac{i}{2} \sum_{I=1}^{\infty}\left(-u_{n,\epsilon_n}^{(I)}(t) \bar u_{m,\epsilon'_m}^{(I)}(t)+\bar{u}_{n,\epsilon_n}^{(I)}(t) u_{m,\epsilon'_m}^{(I)}(t)\right)  = \delta_{nm}\Omega_{\epsilon_n\epsilon'_n}.
\end{equation}
with $\Omega_{\epsilon_n\epsilon'_n}=\left\{U_{n,\epsilon_n}(t), U_{n,\epsilon'_n}(t)\right\}/2$
 the (inverse of the) symplectic form for each mode $n$. }

Now, given two bases with elements ${\bf u}^{(I)}(t)$ and ${\bf w}^{(I)}(t)$, respectively, they will be related by a Bogoliubov transformation with Bogoliubov coefficients  $\alpha_{IJ}$ and $\beta_{IJ}$ as 
\begin{equation}\label{eq:bogou}
    {\bf u}^{(I)}(t) = \sum_{J=1}^{\infty} \alpha_{IJ}{\bf w}^{(J)}(t)+\beta_{IJ}\bar {\bf w}^{(J)}(t).
\end{equation}
such that 
\begin{align}\nonumber
&\alpha_{IJ} = \langle {\bf w}^{(J)}(t), {\bf u}^{(I)}(t)\rangle=\\\nonumber
&\frac{i}{2} \sum_{n=1}^{\infty} {}^{w}\bar{\phi}_{n}^{(J)}(t) {}^{u}\pi_{n}^{(I)}(t)-{}^{w}\bar{\pi}_{n}^{(J)}(t) {}^{u}\phi_{n}^{(I)}(t),\\
&\beta_{IJ} = -\langle \bar {\bf w}^{(J)}(t), {\bf u}^{(I)}(t)\rangle=\\ \nonumber
&-\frac{i}{2} \sum_{n=1}^{\infty} {}^{w}{\phi}_{n}^{(J)}(t) {}^{u}\pi_{n}^{(I)}(t)-{}^{w}{\pi}_{n}^{(J)}(t) {}^{u}\phi_{n}^{(I)}(t).
\end{align}
In consequence, these Bogoliubov coefficients satisfy the conditions
\begin{align}\label{eq:bogo1}
&\sum_{K=1}^{\infty} \alpha_{I K} \bar{\alpha}_{J K}-\beta_{I K} \bar{\beta}_{J K}=\delta_{I J}, \\\label{eq:bogo2}
&\sum_{K=1}^{\infty} \alpha_{I K} \beta_{J K}-\beta_{I K} \alpha_{J K}=0.
\end{align}
The inverse relation to Eq. \eqref{eq:bogou} is given by 
\begin{equation}\label{eq:bogow}
    {\bf w}^{(J)}(t) = \sum_{I=1}^{\infty} \bar\alpha_{IJ}{\bf u}^{(I)}(t)-\beta_{IJ}\bar {\bf u}^{(I)}(t),
\end{equation}
which implies
\begin{align}\label{eq:bogo3}
&\sum_{K=1}^{\infty} \bar\alpha_{K I} {\alpha}_{KJ}-\beta_{KI} \bar{\beta}_{ KJ}=\delta_{I J}, \\\label{eq:bogo4}
&\sum_{K=1}^{\infty} \bar\alpha_{KI} \beta_{ K J}-\beta_{KI} \bar\alpha_{K J}=0.
\end{align}
Besides, if $\bar b_J$ and $b_J$ are creation and annihilation variables in the basis ${\bf w}^{(J)}(t)$, then
\begin{equation}\label{eq:bogob2a}
    b_{J} = \sum_{I=1}^{\infty} \alpha_{IJ}a_{I}+\bar\beta_{IJ}\bar a_{I},
\end{equation}
or equivalently,
\begin{equation}\label{eq:inv-bogo}
    a_{I} = \sum_{J=1}^{\infty} \bar\alpha_{IJ}b_{J}-\bar\beta_{IJ}\bar b_{J}.
\end{equation}
In summary, the Fourier modes $\phi_n$ and $\pi_n$ can be expressed as
\begin{equation}\label{eq:mode-aadagg}
\begin{split}
 &\phi_n(t)=  \sum_{I=1}^\infty a_I u_{n,\epsilon_n=0}^{(I)}(t)+\bar a_I \bar u_{n,\epsilon_n=0}^{(I)}(t),\\ &\pi_n(t)=\sum_{I=1}^\infty a_I u_{n,\epsilon_n=1}^{(I)}(t)+\bar a_I \bar u_{n,\epsilon_n=1}^{(I)}(t).
 \end{split}
\end{equation}

 The quantum field will be written as 
\begin{equation}\label{eq:qmode-aadagg}
\hat\phi_n(t) =  \sum_{I=1}^\infty  u_{n,\epsilon_n=0}^{(I)}(t)\hat a_I+\bar u_{n,\epsilon_n=0}^{(I)}(t)\hat a_I^{\dagger},
\end{equation}
where $(\hat a_I,\hat a_{I'}^{\dagger})$ are the annihilation and creation operators. They satisfy the commutation relations 
\begin{equation}\label{eq:amu-commut}
\left[\hat a_{I}, \hat a^{\dagger}_{I'}\right]=\delta_{II'}\hat{\bf I},\quad \left[\hat a_{I}, \hat {a}_{I'}\right]=0=\left[a^{\dagger}_{I}, a^{\dagger}_{I'}\right],
\end{equation}
for all $I,I'$. Besides, the vacuum state is determined by the condition
\begin{equation}
\hat a_{I}|0\rangle = 0,\quad I=1,2,\ldots\;.
\end{equation}
In our model, particle production at late times will be given by the coefficients $\beta_{IJ}$ of the Bogoliubov transformation between the $in$ and $out$ states bases, the latter determined by the basis of complex solutions $\big({\bf w}^{(I)}(t), \bar {\bf w}^{(I')}(t)\big)$.

\section{Gaussian States and Logarithmic Negativity}\label{sec:gauss-entangl}

Gaussian states are a class of quantum states in continuous variable systems which can be described and completely characterized by two quantities: its mean vector and its covariance matrix \cite{Adesso2014GaussianStatesAndBeyond,adesso2007entanglement}. Therefore, their analysis is computationally straightforward, and several physical quantities, like particle number, quantum entanglement, etc. can be easily derived. Let us introduce the dimensionless field quadrature operators defined from the field operators as

\begin{equation}
    \begin{split}
    \hat{q}_{J}=\frac{1}{\sqrt{2}}(\hat{a}_{J}+\hat{a}_{J}^{\dagger}),\\
    \hat{p}_{J}=\frac{i}{\sqrt{2}}(\hat{a}_{J}^{\dagger}-\hat{a}_{J}).
    \end{split}
    \label{eq: quadratures}
\end{equation}

One can rearrange them in a vector of dimension $2N$, $\hat{\mathbf{R}}=\left( \hat{q}_{1},\hat{p}_1,\dots,\hat{q}_N, \hat{p}_N \right)^{\text{T}}$, and end up with the following commutation relations

\begin{align}
    [\hat{\mathbf{R}}^{i},\hat{\mathbf{R}}^{j}]=i\mathbf{\Omega}^{ij}, && \mathbf{\Omega}\equiv\bigoplus_{N}\begin{pmatrix} 0 & 1 \\ -1 & 0 \end{pmatrix},
    \label{eq: conmutation quadratures}
\end{align}
where $i,j \in \{1,2,...,2N\}$. In the following, if we use lowercase indices they run from $1$ to $2N$, on the contrary, when we use capitalized indices $I,J,...$ they run from $1$ to $N$. $\mathbf{\Omega}$ is a $2N\times2N$ matrix  which is the (inverse of the) symplectic form. The first two moments of this vector, called, respectively, the mean vector and the covariance matrix, are then defined as

\begin{equation}
    \begin{split}
        \boldsymbol{\mu}^{i} \equiv \langle \mathbf{\hat{R}}^{i} \rangle,\\
        \boldsymbol{\sigma}^{ij} \equiv \langle \{ \mathbf{\hat{R}}^{i}-\boldsymbol{\mu}^{i},\mathbf{\hat{R}}^{j}-\boldsymbol{\mu}^{j}\}\rangle,
    \label{eq: covariance matrix}
    \end{split}
\end{equation}
where the expectation value is taken on a quantum state $\hat\rho$ as $\langle{\hat{\cal O}} \rangle={\rm Tr}\left( \mathbf{\hat{\cal O}} \hat{\rho}\right)$. Gaussian states are then defined as the quantum states which can be completely described by these two moments. They are among the best-known states in continuous-variable quantum information, with wide-ranging applications ~\cite{weedbrook2012gaussian}. In our simulations, they will be sufficient to carry out an extensive analysis of entanglement properties.

\jao{\it Elementary Gaussian states: }The vacuum state of an \( N \)-mode system ---i.e., \( |0\rangle_N \) in the Fock basis--- is represented by a zero mean vector and a covariance matrix equal to the \( 2N \times 2N \) identity matrix in our units. This simple structure allows us to initialize the system in a well-defined reference state and track the evolution of entanglement through transformations of the covariance matrix.

Another type of initial state that we will use in our simulations are the one-mode squeezed states. These states arise naturally in quantum optics when a nonlinear medium, such as a crystal, generates photon pairs through a degenerate parametric down-conversion process. The resulting state is a superposition of even photon-number Fock states, and is called ``squeezed'' because one of its quadratures has reduced uncertainty below the (equally distributed) vacuum level, at the expense of increased uncertainty in the conjugate quadrature. They have zero mean vector, while the covariance matrix of a one-mode squeezed vacuum state with squeezing parameter \( r \) is given explicitly by

\begin{equation}
\boldsymbol{\sigma}_{\text{sq}}(r) = 
\begin{pmatrix}
e^{-2r} & 0 \\
0 & e^{2r}
\end{pmatrix}.
\end{equation}

This kind of squeezing can also be interpreted in terms of Bogoliubov transformations. For a given mode \( \hat b \) in Eq. \eqref{eq:bogob2a} satisfying $\hat b = \alpha \hat a+\bar \beta a^\dagger$, i.e. the transformation only mixes each original mode \( \hat a \) with its own conjugate \( \hat a^\dagger \), and does not entangle different modes, the transformation is said to be a single-mode squeezing transformation. In this case, $|\alpha| = \sinh r$ and $|\beta|=\cosh r$. This illustrates how the one-mode squeezing operation---diagonal in the mode basis---can be seen as a special case of more general multimode Bogoliubov transformations, where mode mixing is absent. Note that this Bogoliubov transformation is precisely what induces the squeezing and, in our simulations, it is applied before the boundaries start moving, as a way to prepare the initial state as a nonvacuum state with controlled properties.

Another interesting family of initial states corresponds to two-mode squeezing states. A two-mode squeezer is a Gaussian unitary that entangles two modes by creating correlated photon pairs. Physically, it can be implemented through a nondegenerate parametric down-conversion process.\footnote{It is described by the unitary operator \( S_{2}(r) = \exp[r (\hat{a_1}\hat{a_2} - \hat{a_1}^\dagger \hat{a_2}^\dagger)/2] \), where \( r \) is the squeezing parameter and $\hat a_1$ and $\hat a_2$ are the two involved modes.} 

This transformation is defined as 
\begin{eqnarray}
\hat b_1 &=& \alpha \hat a_1+\bar \beta \hat a_2^\dagger,\\
\hat b_2 &=& \alpha \hat a_2+\bar \beta \hat a_1^\dagger,   
\end{eqnarray}
namely, the transformation mixes the mode \( \hat a_1 \) with \( \hat a_2^\dagger \). Hence, it does entangle different modes. This transformation is said to be a two-mode squeezing transformation (with phase $\phi=0$). In this case, $|\alpha| = \sinh r$ and $|\beta|=\cosh r$. 

When applied to the two-mode vacuum, this transformation produces the two-mode squeezed state, also known as an Einstein-Podolsky-Rosen (EPR) or Bell state, which exhibits perfect correlations in certain quadrature combinations as \( r \to \infty \). The covariance matrix of this state, for squeezing phase \( \phi = 0 \), is given by 
\begin{equation}
\boldsymbol{\sigma}_{\text{EPR}} (r)= \left(\begin{array}{cc}\cosh 2r I_2 & \sinh 2r \sigma_z \\ \sinh 2r \sigma_z & \cosh 2r I_2\end{array}\right),    
\end{equation}
with $I_2$ the $2\times2$ identity matrix and $\sigma_z$ the $2\times2$ Pauli-z matrix. In our simulations, we apply these two-mode squeezing operations in a pairwise manner---mode 1 with mode 2, mode 3 with mode 4, and so on---as a method of preparing entangled initial states with tunable strength.

Finally, since our work is framed within the context of analogue gravity scenarios, it is important to understand how the presence of thermal noise in the initial state affects the entanglement properties of the system. Thermal states are generically defined as those that maximize the von Neumann entropy for a fixed average energy. In bosonic theories, they are Gaussian states fully characterized by a single parameter: the occupation number \( n_{\text{env}} \), which represents the mean number of bosons in a given mode. This number depends on the frequency of the mode and the temperature of the environment according to the Bose--Einstein distribution. Thermal Gaussian states have zero mean vector and a covariance matrix for each mode given by \( \boldsymbol{\sigma} = (2n_{\text{env}} + 1) I_{2 \times 2} \), indicating that thermal noise symmetrically increases the uncertainty in both quadratures. Besides, they are mixed states. 

\jao{\it Gaussian unitaries: }The unitary transformations that takes one Gaussian state to another Gaussian state are called symplectic transformations, as they belong to the symplectic group $Sp(2N,\mathbb{R})$, the set of transformations $\mathbf{S}$ that leave the symplectic form invariant, namely, $\mathbf{S}\mathbf{\Omega}\mathbf{S}^{\text{T}}=\mathbf{\Omega}$. In fact, if the Hamiltonian of the system is, at most, quadratic in the field operators, the time evolution of the state is symplectic and, therefore, the initial state maintains its Gaussian character. In terms of the statistical moments, we define a Gaussian unitary as the transformation 

\begin{align}
    \boldsymbol{\mu}\rightarrow \mathbf{S}\boldsymbol{\mu}+ \boldsymbol{d}, && \boldsymbol{\sigma}\rightarrow \mathbf{S}\boldsymbol{\sigma}\mathbf{S}^{\text{T}}.
\end{align}

If one recalls the relation between the \textit{in} and the \textit{out} modes in the Heisenberg picture given by the Bogoliubov transformations described above after rearranging the field operators in a vector as $\hat{\mathbf{A}}_{(in)} = \left( \hat a_{1}, \hat{a}^\dagger_{1}, ..., \hat a_{N}, \hat{a}^\dagger_{N} \right)$, the Bogoliubov relation in Eq. \eqref{eq:bogob2a} can be arranged as an $\mathbf{S}$-matrix \cite{bhardwaj2024entanglement},

\begin{equation}
    \mathbf{S_{(A)}}=\begin{pmatrix}
        {\alpha}_{11} && \bar{\beta}_{11} && ... && {\alpha}_{N1} && \bar{\beta}_{N1} \\
        \beta_{11} && \bar\alpha_{11} && ... && \beta_{N1} && \bar\alpha_{N1} \\
        ... && ... && ... && ... && ...\\
        {\alpha}_{1N} && \bar{\beta}_{1N} && ... && {\alpha}_{NN} && \bar{\beta}_{NN} \\
        \beta_{1N} && \bar\alpha_{N1} && ... && \beta_{NN} && \bar\alpha_{NN} \\
    \end{pmatrix}.
    \label{eq: S matrix in bogos terms}
\end{equation}\\

Then, the transformation from the $in$ basis to the $out$ one is $\hat{\mathbf{A}}_{(out)} = \mathbf{S_{(A)}} \cdot \hat{\mathbf{A}}_{(in)}$. Note that condition  $\mathbf{S_{(A)}}\mathbf{\Omega}\mathbf{S_{(A)}}^{\intercal}=\mathbf{\Omega}$, namely, that $\mathbf{S_{(A)}}$ belongs to the symplectic group, is equivalent to fulfill the Bogoliubov relations in Eqs. \eqref{eq:bogo1}-\eqref{eq:bogo4}. Moreover, the vector of annihilation and creation variables $\hat{\mathbf{A}}$  can be easily related to the quadrature variables $\hat{\mathbf{R}}$ with the matrix $\mathbf{B}$,
\begin{align}
    \mathbf{\hat{A}} = \mathbf{B}\cdot \mathbf{\hat{R}},&& \mathbf{B} \equiv \bigoplus_{N}\frac{1}{\sqrt{2}}\begin{pmatrix}
        1 && i \\
        1 && -i\\
    \end{pmatrix} .
    \label{eq: transformation matrix to quadratures}
\end{align}
Then, the relation between both $\mathbf{S}$-matrices will be given by $\mathbf{S_{(A)}}= \mathbf{B}\cdot \mathbf{S_{(R)}}\cdot \mathbf{B}^{-1}$. 

Once the $\mathbf{S}$-matrix has been changed to the quadrature basis $\hat{\mathbf{R}}$, it is time to take advantage of the computational benefits that Gaussian states offer. Any covariance matrix of a Gaussian state can be transformed into the so called Williamson's form through a symplectic transformation \cite{williamson1936algebraic}, leaving its shape as

\begin{equation}
    \boldsymbol{\sigma}=\mathbf{S}\boldsymbol{\nu} \mathbf{S}^{\intercal},
\end{equation}
where $\mathbf{S}\in Sp(2N,\mathbb{R})$, and $\boldsymbol \nu$ is the following covariance matrix:
\begin{equation}
    \boldsymbol{\nu} = \bigoplus_{k=1}^{N} \begin{pmatrix}
        v_{k} && 0 \\
        0 && v_{k} 
    \end{pmatrix}.
\end{equation}

The values $v_{k}$ are called the symplectic eigenvalues of the covariance matrix and are invariant under the action of global symplectic transformations of the matrix $\boldsymbol \sigma$. At the same time, due to the commutation relations, any covariance matrix should be \textit{physically realizable}, that is

\begin{equation}
    \boldsymbol{\sigma} + i\boldsymbol{\Omega} \geq 0.
\end{equation}

This latter inequality represents the uncertainty principle of canonical operators in its strong, Robertson-Schr\"odinger form \cite{Adesso2014GaussianStatesAndBeyond}. This actually leads to the following property 
\begin{equation}
    v_{k} \geq 1 \hspace{5pt} \forall k\in\{1,...,N\},
\end{equation}
for the symplectic eigenvalues of any valid covariance matrix.

{\it Entanglement in Gaussian states:} Let us now turn our attention to bipartite quantum entanglement. Here, the positivity of the partial transposition (PPT) criterion plays a central role. It is a necessary and sufficient condition for nonseparability in Hilbert spaces of shape $2\times2$ or $3\times3$. However, in higher dimensional spaces there exist entangled states that scape to this criterion by having a positive partial transpose. Conveniently, in \cite{simon2000peres}, it is shown that the PPT criterion is also necessary and sufficient for all bipartitions of the form $\left( 1 \times (N-1) \right) $ of Gaussian states. In this manuscript, we will restrict our analysis to that kind of bipartition. For general bipartitions of $N_A$ modes in subsystem $A$ and $N_B$ modes in subsystem $B$, such that $ N=N_A+N_B$, under partial transposition the covariance matrix $\sigma_{AB}$ of a bipartite $(N_{A}\times N_{B})$ Gaussian state is transformed into
\begin{equation}
    \boldsymbol{\tilde{\sigma}}_{AB} = T \boldsymbol{\sigma}_{AB} T,
\end{equation}
where $T = I_{2N_{A}} \bigoplus \Sigma_{N_{B}}$ and $\Sigma_{N_{B}} = \bigoplus_{N_{B}}\sigma_z$ is the direct sum of $N_{B}$ matrices of the kind $\sigma_z$, recalling that the latter are $2\times2$ Pauli-z matrices. The PPT criterion shows that a Gaussian state (with $N_{A}=1$ and $N_{B}$ arbitrary) is separable if and only its partially transposed covariance matrix is also a physically realizable covariance matrix. In terms of the symplectic eigenvalues~$\tilde{v}_{k}$ of the partially transposed covariance matrix it means that $\tilde{v}_{k} \geq 1, \forall k \in \{1,N\}$. Then, if one of those symplectic eigenvalues is less than 1, the state is definitely entangled.\footnote{For continuous
variable states, partial transposition of a bipartite state amounts to a reflection (change of sign) of the momentum of only one of the subsystems in the Wigner function representation of the state \cite{simon2000peres}.}

From this criterion an entanglement monotone can be constructed. An entanglement monotone is a functional $E : \mathcal{S}(\mathcal{H}) \longrightarrow \mathbb{R}$, which maps states of the quantum system into real numbers. Such function $E$ is defined by a set of physically motivated properties, as follows \cite{plenio2005logarithmic}:

\begin{enumerate}
    \item $E : \mathcal{S}(\mathcal{H}) \longrightarrow \mathbb{R}$ is a positive functional, and $E(\hat{\rho})=0$ for any separable state $\hat{\rho} \in \mathcal{D}(\mathcal{H})$.
    \item $E$ does not increase on average under Local Operations and Classical Communication (LOCC), that is, if in a LOCC (or PPT) protocol applied to the state $\hat{\rho}$, the state $\hat{\rho}_{i}$ is obtained with probability $p_{i}$, then,
    \begin{equation}
        E(\hat{\rho})\geq \sum_{i=1}p_{i}E(\hat{\rho}_{i}).
    \end{equation} 
\end{enumerate}

The logarithmic negativity is especially relevant in these scenarios. For a bipartition in subsystems $A$ and $B$, it is defined generally as

\begin{equation}
    {\rm LogNeg}(\hat{\rho}_{AB}) \equiv \log\parallel\hat{\rho}_{AB}^{T_{B}}\parallel_{1}= \log(2\mathcal{N}(\hat{\rho}_{AB})+1), 
\end{equation}
where $\hat{\rho}_{AB}$ is the initial bipartite state, $T_{B}$ means partial transpose with respect to subsystem $B$, $\parallel\cdot\parallel_{1}$ denotes the trace norm (which is the sum of the singular values or absolute eigenvalues of the operator), and $\mathcal{N}$ is the negativity of the state (see \cite{vidal2002computable}). In the case of Gaussian states, it only depends on the covariance matrix. It is simply given by

\begin{equation}
    {\rm LogNeg}(\boldsymbol{\sigma}_{AB}) = \sum_{k=1}^{N} \max[0, -\log{\Tilde{v}_{k}}].
\end{equation}

The logarithmic negativity has been shown to be an upper bound to the distillable entanglement \cite{vidal2002computable} as well as a measure of the PPT-entanglement cost for some quantum states \cite{audenaert2003entanglement}. However, it does not have a direct physical meaning as other entanglement monotones and measures do. Despite this, logarithmic negativity is a strong candidate for measuring entanglement in Gaussian states. This is mainly because, since it is derived directly from the PPT criterion, a nonzero value of $\rm LogNeg$ is both necessary and sufficient to detect bipartite entanglement in Gaussian states for bipartitions of the $1\times (N-1)$ form. Not only that, but due to its definition, it is also easily computable, avoiding minimizations over infinite sets, as is the case with many other entanglement measures.

Another interesting property of $\rm LogNeg$ is that, for separable states, it satisfies $ {\rm  LogNeg}(\rho_1 \otimes \rho_2 ) = {\rm  LogNeg} ( \rho_1) + {\rm  LogNeg}(\rho_2)$, allowing one to study the correlations between the subsystems of each $\rho_i$ independently, without interference.

Finally, the main drawback of logarithmic negativity is that it does not satisfy the monogamy relation,

\begin{equation}
    E_{A|BC} \geq E_{A|B} + E_{A|C}, 
\end{equation}
where $E$ is an entanglement measure (e.g., ${\rm  LogNeg}$). This implies that, although ${\rm  LogNeg}$ is excellent for studying bipartite entanglement between subsystems of a Gaussian state, the way in which entanglement is distributed within the subsystem is not fully captured by this monotone. Hence,  some sort of minimization over infinite sets is required \cite{Gao2021Monogamy}.

\section{The Hawking partner}\label{sec:partner}

So far we have computed quantum entanglement focusing on bipartitions of the form $1\times(N-1)$. The aim was to determine the structure of these quantum correlations of a given mode within the thermal frequency band (a Hawking mode) with the rest of the system. But we also know from black hole physics that each Hawking mode is maximally entangled with its partner. Hence, it will be enlightening to construct the Hawking partner of a given Hawking mode, compute the quantum entanglement between these two modes and compare the entanglement structure for this kind of bipartition of the system.

In order to construct the Hawking partner of a given Hawking mode, we will follow the procedure in Ref. \cite{hotta2015Partner} by Hotta-Sch\"utzhold-Unruh and denote it as the HSU formula in what follows. There, the partner mode is constructed out of the Hawking mode, choosing a reference vacuum state. \jao{In Ref. \cite{hotta2015Partner} there are several conditions required for the construction. The first one, there denoted by condition {\bf A}, requires} that the reduced density matrix of the Hawking and the partner modes obtained by integrating out all other degrees of freedom should be a \jao{pure state, in the sense that it must have zero entanglement with the rest of the system. This condition is automatically fulfilled in our setting by virtue of {\it Theorem~1} of Ref. \cite{Botero:2003wov}, which states that any pure Gaussian state can always be decomposed as a product state of entangled two-mode squeezed states and single-mode vacuum states. However, the HSU formula yields partner modes that are not commuting, limiting its use to describe the system solely in terms of pairs of Hawking-partner modes, without further considerations. Nevertheless, as we will see, this formula has very interesting applications}. Besides, in order to uniquely fix the partner mode, there is another condition that must be satisfied. Concretely, the partner mode will be constructed such that the quantum state that results from annihilating one partner particle should be the same state after creating one Hawking particle (this is criterion {\bf B1} in Ref. \cite{hotta2015Partner}). There is an alternative possibility (criterion {\bf B2} in Ref. \cite{hotta2015Partner}) where the quantum state after annihilating  one Hawking particle is the same
state after creating one partner particle. However, we have checked that this \jao{last} possibility cannot be \jao{generically} realized in our setting. We will explain why below. 

This construction results very clear if we work with annihilation and creation operators. From the perspective of $in$ observers, they measure the field on the $in$ basis $(\hat a_I,\hat a_I^\dagger)$. They realize that the field, once the cavity becomes at rest, is not in the vacuum anymore but, rather, in an excited state. In fact, they can identify a particular Hawking mode $b_J$ (in the appropriate frequency band) such that it can be expressed in the $in$ basis as in Eq.~\eqref{eq:bogob2a}. In the following, let us fix $J$ within the thermal frequency band, and let us also refer to this particular mode as $\hat{b}_H$. Then, this mode takes the form
\begin{equation}\label{eq:bh-aI}
   \hat b_{H} = \sum_{I=1}^{N} {\alpha}_I \hat a_I+\bar{\beta}_I \hat{a}_I^{\dagger},
\end{equation}
where, for simplicity, we have omitted the second label $J$ in the Bogoliubov coefficients ---see again Eq.~\eqref{eq:bogob2a}. 

If we rearrange all annihilation operators $\hat{a}_J$ into a vector $\hat{\mathbf{a}}$, and similarly for the Bogoliubov coefficients, we can redefine the Hawking mode as

\begin{equation}
    \hat b_H = \langle \boldsymbol{{\bar\alpha}} | \hat{\mathbf{a}} \rangle + \langle \hat{\mathbf{a} }| \boldsymbol{\bar{\beta}} \rangle,
\end{equation}
such that the first addend is obviously defined as 
\begin{equation}
    \langle \boldsymbol{{\bar\alpha}} | \hat{\mathbf{a}} \rangle = \sum_{I=1}^{N} {\alpha}_I \hat a_I,
\end{equation}
and similarly for the second addend. We can define the orthonormal vectors $\mathbf{n}_{\parallel}$ and $\mathbf{n}_{
\perp}$ given by the subspace spanned by $\boldsymbol{\alpha}$ and $\boldsymbol{\beta}$, in such a way that
\begin{equation}
    \boldsymbol{\alpha} = \alpha \mathbf{n}_{\parallel}, \quad \boldsymbol{{\beta}} = {\beta}_{\parallel}\mathbf{n}_{\parallel} + {\beta}_{\perp} \mathbf{n}_{\perp},
\end{equation}
with $\alpha =\langle \mathbf{n}_{\parallel} | \boldsymbol{{\alpha}} \rangle$, and $\beta_{\parallel}=\langle \mathbf{n}_{\parallel} | \boldsymbol{{\beta}}\rangle$ and $\beta_{\perp}=\langle \mathbf{n}_{\perp} | \boldsymbol{{\beta}}\rangle$. Let us also define the annihilation modes 
\begin{equation}
\hat a_{\parallel} = \langle \mathbf{n}_{\parallel} | \hat{\mathbf{a}}\rangle, \quad  \hat a_{\perp} = \langle \mathbf{n}_{\perp} | \hat{\mathbf{a} }\rangle,
\end{equation}
which satisfy the usual commutation relations. With this, we can write the Hawking mode as
\begin{equation}\label{eq:bh-mode}
    \hat b_{H} = {\alpha} \,\hat a_{\parallel} +  \bar{\beta}_{\parallel} \hat a_{\parallel} ^{\dagger}+ \bar{\beta}_{\perp} \hat a_{\perp} ^{\dagger}.
\end{equation}
Note that $\beta_{\perp} = 0$ implies that the Hawking mode is a one-mode squeezing and there is no need for a partner particle. Otherwise, 
following \cite{hotta2015Partner}, 
in order to satisfy condition \jao{{\bf A}}, 
the partner mode must be of the form
\begin{equation}\label{eq:bp-mode}
    \hat b_{P} = {\gamma}_{\parallel} \hat a_{\parallel} + {\gamma}_{\perp} \hat a_{\perp} + \bar{\delta}_{\parallel} \hat a_{\parallel}^{\dagger} + \bar{\delta}_{\perp} \hat a_{\perp}^{\dagger}.
\end{equation}

Since $\hat b_H$ and $\hat b_P$ must obey the usual commutation relations, the previous parameters must satisfy: 
\begin{align}\label{eq:hp-cond1}
    |\alpha|^2 - |\beta_{\parallel}|^2 - |\beta_{\perp}|^2 = 1, \\ \label{eq:hp-cond2}
    |\gamma_{\parallel}|^2+ |\gamma_{\perp}|^2 - |\delta_{\parallel}|^2- |\delta_{\perp}|^2 = 1, \\ \label{eq:hp-cond3}
    \bar{\gamma}_{\parallel} \alpha = \bar\beta_\parallel {\delta}_\parallel + \bar\beta_\perp {\delta}_\perp, \\ \label{eq:hp-cond4}
    \bar{\alpha} {\delta}_\parallel = \bar{\gamma}_\parallel {\beta}_\parallel + \bar{\gamma}_\perp {\beta}_\perp.
\end{align}
The first condition is fulfilled as a consequence of Eq.~\eqref{eq:bogo1}. The last three conditions do not uniquely specify the partner mode. Here, we need to introduce condition {\bf B1} of Ref. \cite{hotta2015Partner}. It amounts to $\delta_\perp = 0$, and allows us to fix the remaining  three coefficients up to an irrelevant global phase. 

\jao{In our case, we have in general $\beta_\parallel \neq 0$, but their value is negligible except for the smallest $out$ frequencies $\omega_J$ (see Sec. \ref{sec:numerics}). When the single-mode squeezing approximately vanishes ($\beta_\parallel \simeq 0$ and $\beta_\perp =\beta$), the Hawking mode and the partner mode are related to the $in$ modes via a two-mode squeezing transformation, namely,
\begin{equation}\label{eq:HP-2mode-squeez}
  \hat b_H \simeq \alpha \,\hat a_{\parallel} +  \bar{\beta} \hat a_{\perp} ^{\dagger},\quad b_P \simeq \alpha \hat{a}_{\perp} + \bar{\beta} \hat{a}_{\parallel}^{\dagger}.
\end{equation}
}

We would like to briefly comment on the criterion {\bf B2}. Its implementation requires that $\beta_\parallel$ be small enough. As one can see below, for Hawking modes in the limit $J\gg 1$, we see that $\beta_\parallel$ decreases very fast. However, we have checked that imposing $\boldsymbol{\gamma}\!\!\parallel\!\! \boldsymbol{\beta}$ does not allow us to consistently solve Eqs. \eqref{eq:hp-cond1}-\eqref{eq:hp-cond4}, even in the limit $J\gg 1$ for almost none of the $out$ modes $J$ (in the case of large accelerations of the boundary, only a few modes satisfy condition {\bf B2}). In summary, $\beta_\parallel$ is barely small enough for criterion {\bf B2} to be applicable in our case. In what follows, we will adhere to criterion {\bf B1}, unless otherwise specified. Consequently, all claims will consistently depend on this choice.

In general, the partner mode can also be written as the Hawking mode in Eq. \eqref{eq:bh-aI}, namely,
\begin{equation}\label{eq:bp-aI}
   \hat b_{P} = \sum_{I=1}^{N} {\gamma}_I \hat a_I+\bar{\delta}_I \hat{a}_I^{\dagger},
\end{equation}
where $\gamma_{I}$ and $\delta_{I}$ are the components of the vectors $\boldsymbol{\gamma}=\gamma_{\parallel}{\bf n}_{\parallel}+\gamma_{\perp}{\bf n}_{\perp}$ and $\boldsymbol{\delta}=\delta_{\parallel}{\bf n}_{\parallel}$, respectively.

An interesting advantage of the HSU formula \jao{for pure states} is the simplicity and efficiency in calculating the entanglement entropy (logarithmic negativity). Just by looking at Eqs. \eqref{eq:bh-mode} and \eqref{eq:bp-mode}, one quickly realizes that we need to deal with two (rather than $N$) modes, and all the information encoded in the vectors $\alpha_I$ and $\beta_I$ can actually be compressed into $\alpha$, $\beta_\parallel$ and $\beta_\perp$. As we will see in detail below, it tremendously reduces computational times. At the same time, it fully agrees with the logarithmic negativity for a bipartition with subsystem $A$ as the corresponding Hawking mode, and subsystem $B$ for the rest of the system. Hence, we do not need to carry out a full tomography of the quantum state, but only knowledge of the parameters $\alpha$, $\beta_\parallel$, and $\beta_\perp$. 

We want to remark on several aspects of the HSU formula (see Appendix \ref{app:HSU-alg} for more details). On one hand, we can only apply it to pure states. For mixed states (like thermal ones) it is conceptually unclear, or at least ambiguous, what must be purified. Hence, for initial thermal states, we do not adopt this method in order to compute the logarithmic negativity of a Hawking mode and its partner, but instead the $1\times(N-1)$ logarithmic negativity. On the other hand, if we restore the index $J$ corresponding to each $out$ mode in the annihilation variables as $b_{H,J}$, we can compute the corresponding $\hat a_{\parallel,J}$ and $\hat a_{\perp,J}$ for each $J$. One can see that these variables satisfy the usual commutation relations, except when considering different $out$ modes. Concretely, for two Hawking modes $J$ and $J'$, all commutators of the corresponding parallel and perpendicular annihilation and creation operators are the usual ones, except for
\begin{eqnarray}\nonumber
    &&[\hat a_{\parallel,J},\hat a_{\parallel,J'}^\dagger] = \langle {\boldsymbol{n}}_{\parallel,J'} | \boldsymbol{n}_{\parallel,J}  \rangle,\quad [\hat a_{\parallel,J},\hat a_{\perp,J'}^\dagger] = \langle {\boldsymbol{n}}_{\perp,J'} | \boldsymbol{n}_{\parallel,J}  \rangle,\\
    &&[\hat a_{\perp,J},\hat a_{\parallel,J'}^\dagger] = \langle {\boldsymbol{n}}_{\parallel,J'} | \boldsymbol{n}_{\perp,J}  \rangle,\quad [\hat a_{\perp,J},\hat a_{\perp,J'}^\dagger] = \langle {\boldsymbol{n}}_{\perp,J'} | \boldsymbol{n}_{\perp,J}  \rangle.\nonumber
\end{eqnarray}
In general, they are not vanishing. In addition, in our simulations, the above inner products can be of the order of the unit. As a consequence, despite $\hat b_{H,J}$ and $\hat b^\dagger_{H,J}$ satisfying the usual commutation relations, one can see that
\begin{equation}\label{eq:comm-bp-bh}
[\hat b_{P,J}, \hat b_{H,J'}]=\sum_I\gamma_{IJ}\bar\beta_{IJ'}-\bar\delta_{IJ}\alpha_{IJ'},    
\end{equation}
is not guaranteed to always vanish, among other commutators, with $\hat b_{P,J}$ being the annihilation operators of each partner mode, and
$\gamma_{IJ}$ and $\delta_{IJ}$ being the respective components of the vectors,
\begin{equation}\label{eq:partners-J}
    \boldsymbol{\gamma}_{J}=\gamma_{\parallel,J}{\bf n}_{\parallel,J}+\gamma_{\perp,J}{\bf n}_{\perp,J},\quad \boldsymbol{\delta}_{J}=\delta_{\parallel,J}{\bf n}_{\parallel,J}.
\end{equation}
We have numerically checked that these commutators do not vanish in general (see Appendix~\ref{app:HSU-alg}). One could argue that the commutators are not symplectic-independent quantities. However, one can easily realize that they cannot be made zero altogether by a symplectic transformation.\footnote{In Ref. \cite{Osawa:2025dja} it is possible to find an example where symplectic invariants are given as combinations of those commutators.} This is a mere consequence of the nontrivial multimode entanglement structure of the final state of the system.  In general, two Hawking modes $J$ and $J'$ have nonvanishing quantum entanglement ---see Appendix \ref{app:one-by-one-logneg}.  In summary, the HSU formula cannot be used to describe the entire system as a canonical basis of pairs of Hawking-partner modes, unless all these commutators vanish for all $J\neq J'$, namely, Hawking and partner modes for  $J\neq J'$ must be uncorrelated. In our setting, these correlations do not vanish (see Appendix \ref{app:HSU-alg}). \jao{On the other hand, we do expect that, whenever the final state is exactly thermal, Hawking modes will be uncorrelated. However, in our case, this would correspond to an evolution from a pure state to a mixed state, breaking unitarity. Nevertheless, if the evolution is unitary, the final state will be pure and Hawking modes must be correlated, as it is in our case. Let us mention that for trajectories where the acceleration lasts for very long times, we do expect that correlations between Hawking modes will be suppressed. For a general discussion on this issue, we refer the readers to Ref.~\cite{Aurell:2023wqa}.}


\section{Numerical simulations for an expanding cavity}\label{sec:numerics}

In this section, we will study the entanglement entropy (${\rm  LogNeg}$) \jao{in the quantum field resulting from several configurations of the boundaries of the cavity}, how it is distributed between different modes, how it is also encoded in the basis of Hawking-partner pairs, and \jao{its} response regarding some initial squeezing and its robustness against thermal noise. \jao{Specifically, we} will consider a configuration of the cavity where the boundaries follow the trajectory
\begin{align}\nonumber
   f(t)= & 0, \\ \nonumber
   g(t)= & 1 +\frac{s}{2\kappa}+\frac{1}{2\kappa}\bigg[\log\Big(\cosh\big(\kappa(t-t_0)\big)\Big)\notag\\
   -&\log\Big(\cosh\big(s-\kappa(t-t_0)\big)\Big)\bigg].
   \label{eq:Traj-1plt}
\end{align}
Namely, one of the boundaries remains at rest while the other accelerates and stops, symmetrically, at a final position. \jao{This trajectory will represent the formation of a black hole horizon that turns into a white hole horizon with the same surface gravity, which eventually evaporates completely, leaving no remnant.} The dynamics and particle production of this setting have already been analyzed \jao{in some detail in Ref. \cite{GarciaMartin-Caro:2023jjq}. A deeper analysis, including more general configurations, can be found in Ref. \cite{GarciaMartin-Caro:2025xgr}.} In this \jao{particular configuration \eqref{eq:Traj-1plt}}, at times  $t \ll t_0$, with $t_0$ being some finite time, the left boundary remains at position $x^f=0$ at all times, while the right boundary is nearly static at initial position $x^g_{in}=1$ \jao{(recall that we have set $L_0=1$ for convenience)}. Then, at $t\gg T\simeq t_0+\epsilon$ (with $\epsilon=s/\kappa$) its final position will be $x^g_{out} =(1+\epsilon)$, and it will remain at rest at all later times. In the interval $[t_0,t_0+\epsilon]$ the right boundary follows an acceleration, \jao{first positive and then negative}, reaching speeds close to the speed of light. \jao{A trajectory with large acceleration means the boundary reaches the speed of light very quickly compared with the inverse of the fundamental frequency of the quantum field, in this case $\omega_1=\pi/L_0$. In this respect, we must note that at early times, the dispersion relation is given by $\omega_n = \frac{n\pi}{L_0}$, as one can see from Eq. \eqref{eq:RnmSnm-b}, where $n=1,2,\ldots$ is a positive integer. More importantly, given the natural vacuum state in this past asymptotic region, the label $I=1,2,3,\ldots$ associated with the annihilation variables that define the $in$ vacuum can be naturally associated with the label $n$ of the $in$ frequencies. In the following, we will refer to the $in$ frequencies as $\omega_I = \frac{I\pi}{L_0}$ (we have restored $L_0$ for convenience). Their gap will then be $\Delta\omega_I=\pi/ L_0$. We can make the same identification with the $out$ frequencies $\omega_m = \frac{m\pi}{L_0(1+\epsilon)}$ with $m=1,2,3,\ldots$ Again, the labels $J$ of the annihilation variables defining the $out$ vacuum are in one-to-one relation with the label $m$ in the asymptotic future. Hence, at late times, the $out$ frequencies of the field will be given by $\omega_J= \pi J/ (L_0(1+\epsilon))$ and their gap by $\Delta\omega_J=\pi/ (L_0(1+\epsilon))$. }

\jao{The trajectories of the boundaries above} are known to excite the $in$ vacuum into a state at late times where the infrared modes show a nearly thermal spectrum of \jao{particles, with respect to the $out$ vacuum,} dictated by the (modified) Fulling-Davies spectrum \cite{GarciaMartin-Caro:2023jjq,GarciaMartin-Caro:2025xgr}
\begin{equation}\label{eq:betas-fit}
|\beta^{\text{(f)}}_{IJ}|^2 = \frac{2\Delta \omega_I\Delta \omega_J}{\pi\kappa\omega_I}\frac{\Gamma_\beta (\epsilon,\omega_J)}{(e^{2\pi\omega_J/\kappa}-1)},
\end{equation}
\jao{where}
\begin{equation}\label{eq:Gamma}
\Gamma_\beta(\epsilon,\omega_J)=\left[A_\beta+B_\beta\sin^2\big({\cal T}_\beta \, \omega_J\big)\right],
\end{equation}
is a greybody factor that depends strongly on both $\epsilon$ and the $out$ frequencies. It captures the effects of the finite duration of the acceleration,\footnote{Greybody factors with an oscillator behavior are common for trajectories where the mirrors do not move forever. See, for instance, the $\beta$-coefficients in Eq. (7) of Ref. \cite{Good:2019tnf}.} leading to oscillations that are well approximated by a sinusoidal function with a characteristic period given by ${\cal T}_\beta = (1 + C_\beta)\epsilon$. Here, $C_\beta$ is a small, dimensionless parameter with only a weak dependence on the trajectory parameters and on the $in$ and $out$ frequencies. The coefficients $A_\beta$ and $B_\beta$ are also dimensionless, where $A_\beta$ is small and $B_\beta$ is of order 1 --- both exhibit weak dependence on the same set of parameters. This parametrization is valid only for mode sets that have enough time to thermalize at a temperature determined by the acceleration of the boundary and that are not contaminated by the transients, which in turn depends on the specific boundary trajectories. A perfectly thermal state requires that $\Gamma_\beta = 1$. Let us note that the recent analysis of Ref. \cite{GarciaMartin-Caro:2025xgr} shows that the thermal spectrum \jao{(in the infrared sector)} is quite robust if the boundary goes back to its initial position following a time-symmetric trajectory or a transient with smaller acceleration than the expansion. Moreover, if one repeats this process once or twice, the infrared part of the spectrum retains its thermal character. For concrete details see Ref. \cite{GarciaMartin-Caro:2025xgr}. Here, we will focus on particular configurations already considered there, corresponding to relatively small and large values of $\kappa$, keeping $\epsilon$ smaller than unit (as it would be required by recent experiments~\cite{Wilson_2011,Johansson:2009zz,Johansson:2010vqd,Doukas:2014bja,Pasi2013}).

\subsection{Small accelerations of the boundary}

Let us start with relatively small accelerations $\kappa$ of the boundary. In all cases, qualitatively similar results have been found. Consider a concrete realization given by $\epsilon=0.375$ and $\kappa=33.3$. 

\begin{figure}[H]
{\centering     
  \includegraphics[width = 0.48\textwidth]{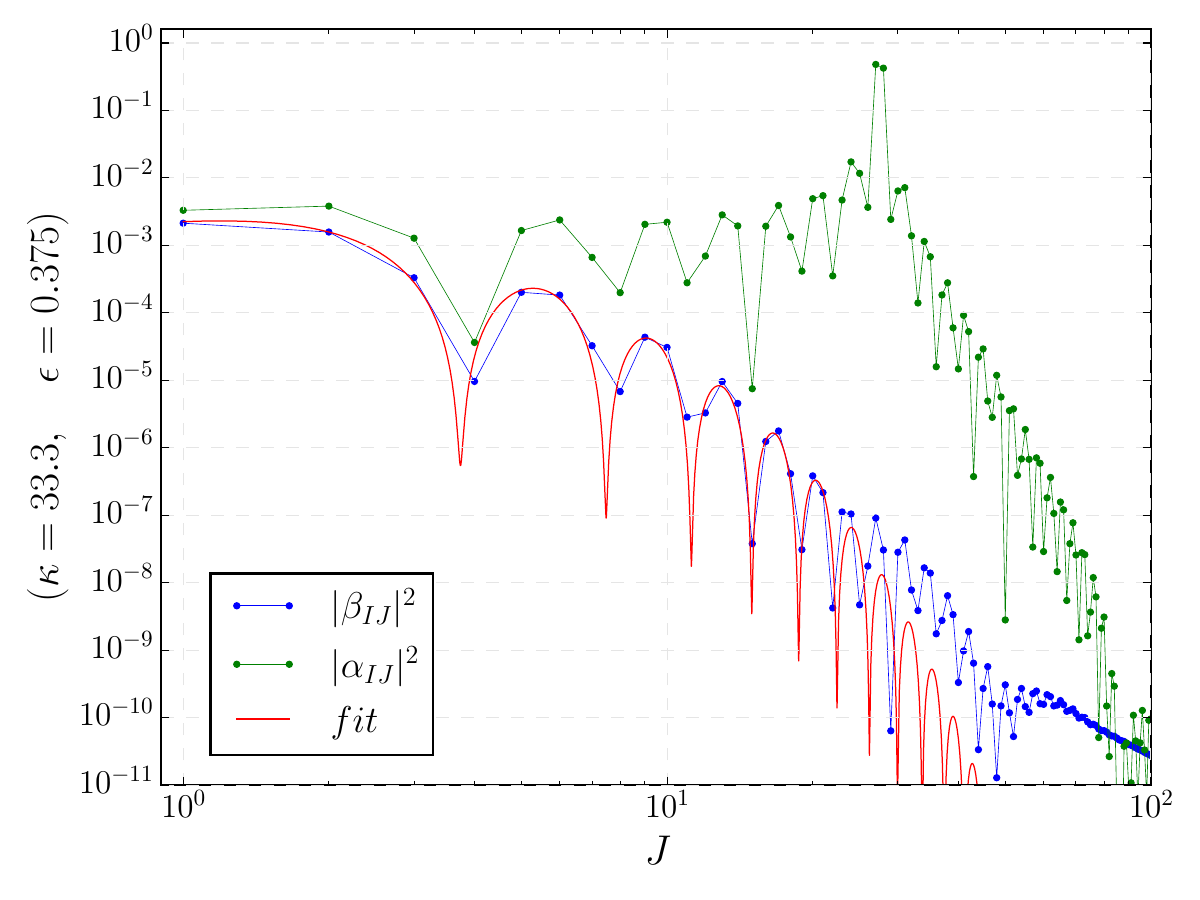}
  \includegraphics[width = 0.48\textwidth]{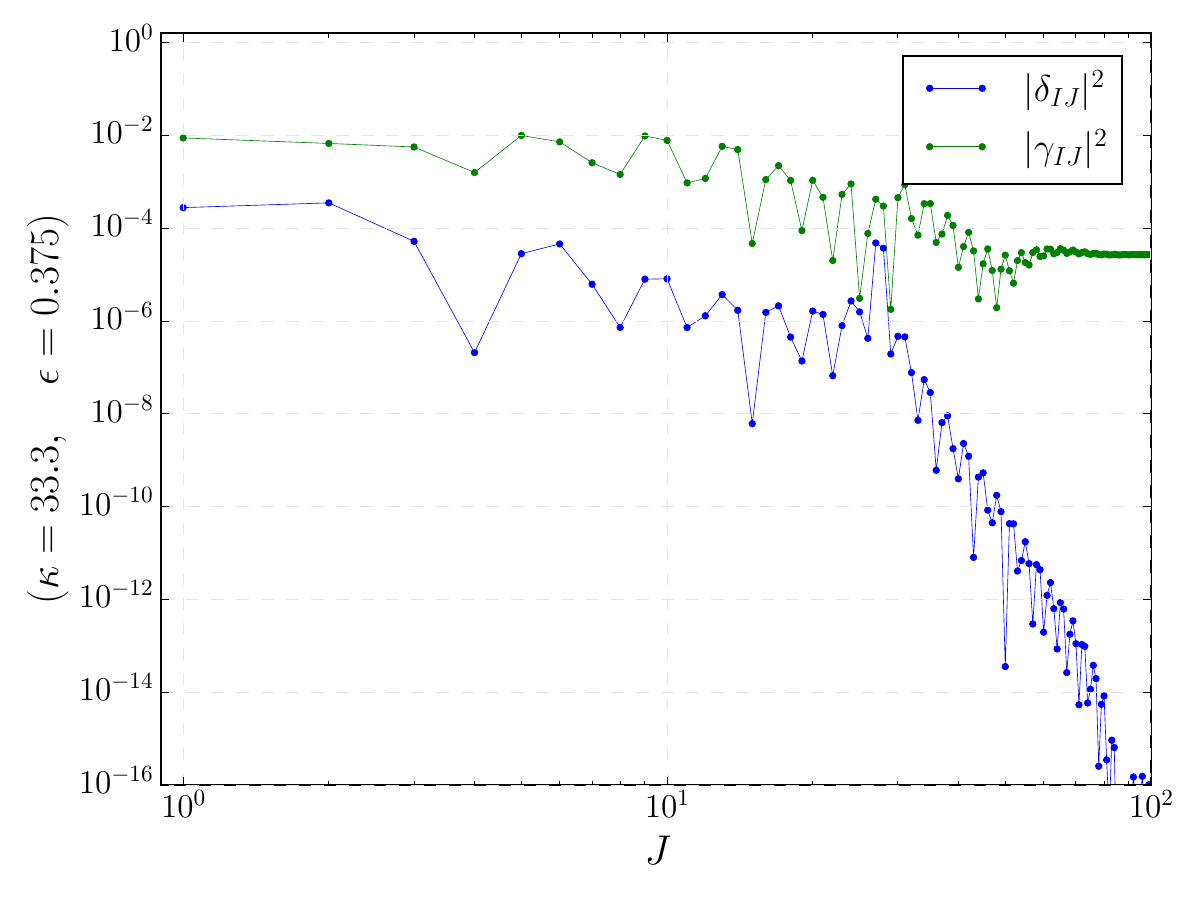}
}
\caption{\justifying Hawking modes and their partners: these plots correspond to the trajectories given by Eq. \eqref{eq:Traj-1plt}, with $\epsilon=0.375$ and $\kappa=33.3$. We show the (modulus squared of the) Bogoliubov coefficients of a Hawking mode (upper panel) and the ones of its partner (lower panel) for a fixed  $in$ frequency $I=20$ and $N\to\infty$ via Richardson extrapolation of the simulations with $N=256,\,512,\,1024$, and as functions of the $out$ frequencies in the interval $[1,100]$. }
\label{fig:1plt-HP}
\end{figure}

\jao{\it Bogoliubov coefficients: } In Fig. \ref{fig:1plt-HP} we show the Bogoliubov coefficients of the Hawking modes for a characteristic $in$ frequency (in this case $I=20$) within the thermal frequency band (upper panel) and those of the corresponding partners (lower panel), both in the limit $N\to\infty$ (Richardson extrapolation out of the simulations with total numbers of modes equal $N=256,512, 1024$).

We also include in the upper panel the fitting expression in Eq. \eqref{eq:betas-fit}. As we see, the behavior of the Bogoliubov coefficients of Hawking modes and corresponding partners is rather different. The $\beta$-coefficients of the Hawking modes at low frequencies agree very well with the fitting expression. At high frequencies, they decay following a power law. On the other hand, the $\alpha$-coefficients at low frequencies oscillate around a constant value (in log scale), reach a peak around the $in$ mode (in this case the peak is around $I\simeq 30$), and then it sharply decreases until it reaches the most ultraviolet frequencies, where it also decays with the $out$ frequency following a power law (not shown in the plot). 

\begin{figure}[H]
{\centering     
\includegraphics[width = 0.48\textwidth]{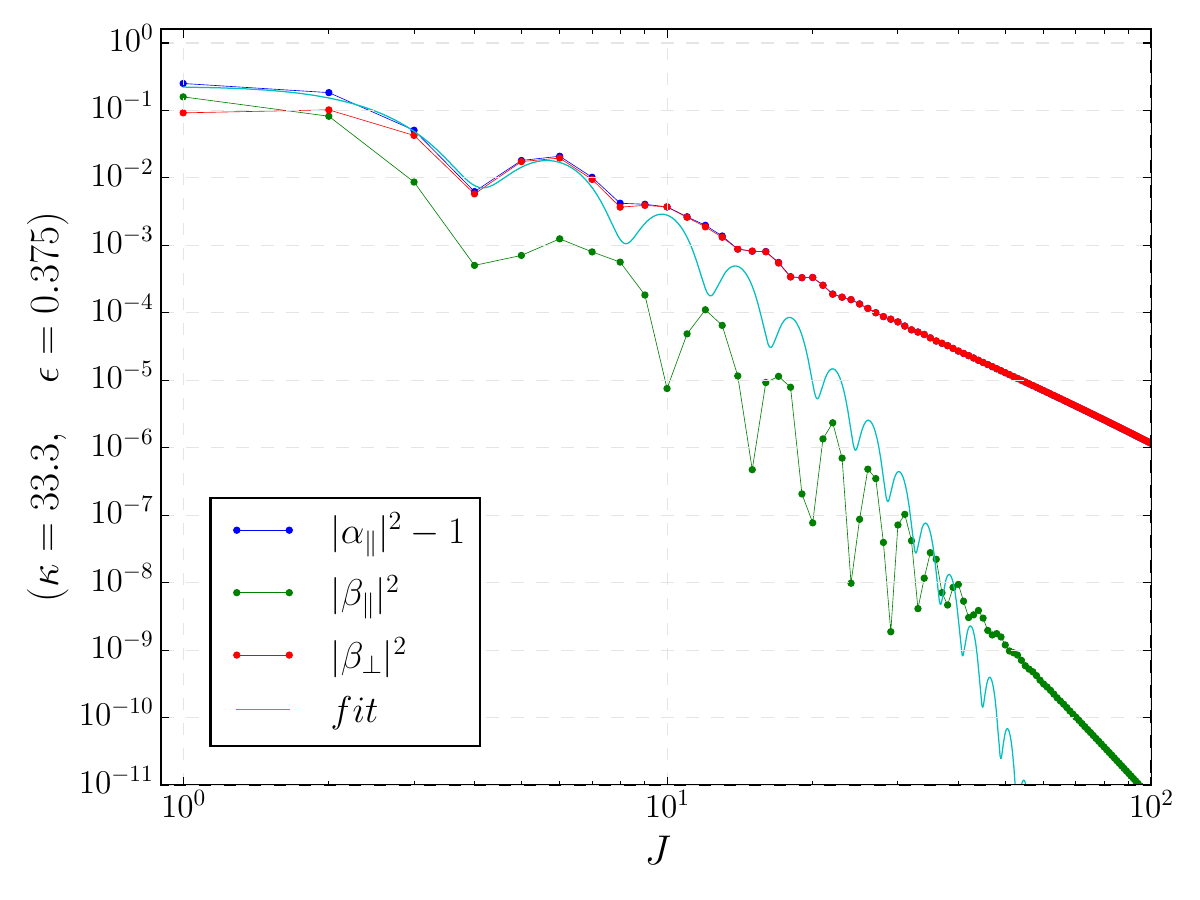}
  \includegraphics[width = 0.48\textwidth]{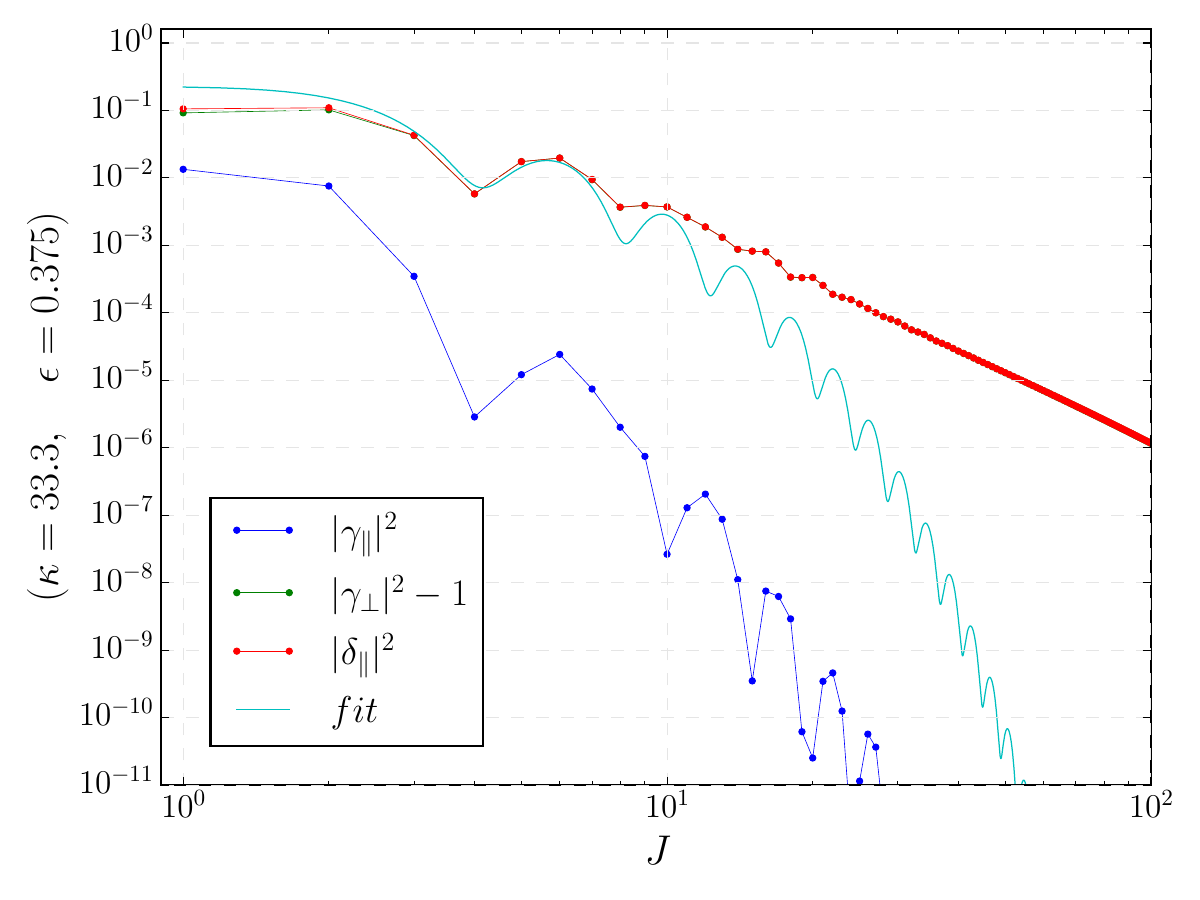}
}
\caption{\justifying Hawking modes and their partners: these plots correspond to the trajectories given by Eq. \eqref{eq:Traj-1plt}, with $\epsilon=0.375$ and $\kappa=33.3$. The upper panel shows the Bogoliubov coefficients in Eq. \eqref{eq:bh-mode} of all Hawking modes and the lower panel those in Eq. \eqref{eq:bp-mode},  for the corresponding partner modes as functions of the  $out$ frequencies in the interval $[1,100]$ and in the limit $N\to\infty$ (via Richardson extrapolation of $N=256,512,1024$). Besides, we include in both cases the fitting expression in Eq.~\eqref{eq:betas-fit2}.}
\label{fig:1plt-HP-B}
\end{figure}

Regarding the partner modes, we also show the behavior of their Bogoliubov coefficients as functions of the $out$ frequencies, defined in Eq. \eqref{eq:partners-J}. Concretely, for the $in$ frequency $I=20$, the $\gamma$-coefficients oscillate around a constant value (in log scale), they slightly decay for intermediate frequencies around $I=30$ and reach a constant value in the ultraviolet sector. Besides, its $\delta$-coefficients also oscillate in the infrared frequency band, with a red tilt, then they reach a peak around $I=30$, close to the $in$ mode $I=20$, and then they sharply decay toward the ultraviolet sector, faster than the Bogoliubov coefficients of the Hawking modes. \jao{Other $in$ modes $I\neq 20$ within the thermal band show a qualitatively similar behavior, hence we do not show them here. } 

In summary, the partner modes have $\gamma$-coefficients that are nearly constant at all $out$ frequencies (with an amplitude of 2 orders of magnitude of difference at infrared and ultraviolet frequencies, respectively), and $\delta$-coefficients are nearly constant at infrared $out$ frequencies with a peak around the frequency of the Hawking mode and a sharp suppression in the ultraviolet sector.
This last behavior indicates that there will be almost no high-energetic particles associated with the partner modes. Hence, we can safely conclude that purification of Hawking modes is a low-energy process. We must recall that we have adopted criterion {\bf B1} in order to fix the Bogoliubov coefficients of the partner modes and for the particular family of symmetric trajectories of the boundaries considered here. Other choices might yield different results.\footnote{\jao{Actually, we do expect that, if the mirror sharply stops, there will be high energetic modes involved.} }

In Fig. \ref{fig:1plt-HP-B} we also show the Bogoliubov coefficients in Eqs. \eqref{eq:bh-mode} and \eqref{eq:bp-mode} as functions of the $out$ frequencies, recalling that we always set $\delta_{\perp} = 0$. 
We see that $|\alpha_J|^2-1\simeq|\beta_{\perp,J}|^2 $, with $|\beta_{\parallel,J}|$ negligible, for Hawking modes, except for the most infrared ones, and $|\gamma_{\perp,J}^2|-1\simeq|\delta_{\parallel,J}|^2 $, with $|\gamma_{\parallel,J}|$ negligible, for all Hawking partners. Hence, we will have $|\gamma_{\perp,J}^2|\simeq |\alpha_{J}^2|$ and $|\delta_{\parallel,J}|^2\simeq |\beta_{\perp,J}|^2$. On one hand, this indicates that, in those cases where these approximations hold, each Hawking-partner pair behaves as a two-mode squeezed state ---see Eq. \eqref{eq:HP-2mode-squeez} and Sec. \ref{sec:gauss-entangl}. Therefore, we conclude that an expanding cavity acts as a squeezing device regarding a given Hawking ($out$) mode and its partner (the mode that purifies the former). On the other hand, we also see that for the most infrared modes, $|\beta_{\perp,J}|^2$ and $|\delta_{\parallel,J}|^2$ agree quite well with the fitting expression
\begin{equation}\label{eq:betas-fit2}
|\beta^{\text{(f)}}_{J}|^2 = \frac{8\Delta \omega_J}{\pi\kappa}\frac{\Gamma_\beta (\epsilon,\omega_J)}{(e^{2\pi\omega_J/\kappa}-1)},
\end{equation}
with the same greybody factor as given in Eq. \eqref{eq:Gamma}.

We have also studied whether criterion {\bf B2} (i.e., $\boldsymbol{\gamma}\!\!\parallel\!\! \boldsymbol{\beta}$) is applicable here, and the answer is in the negative: we have not found consistent solutions to Eqs. \eqref{eq:hp-cond1}-\eqref{eq:hp-cond4} in this case.

\begin{figure}[ht]
{\centering     
\includegraphics[width = 0.48\textwidth]{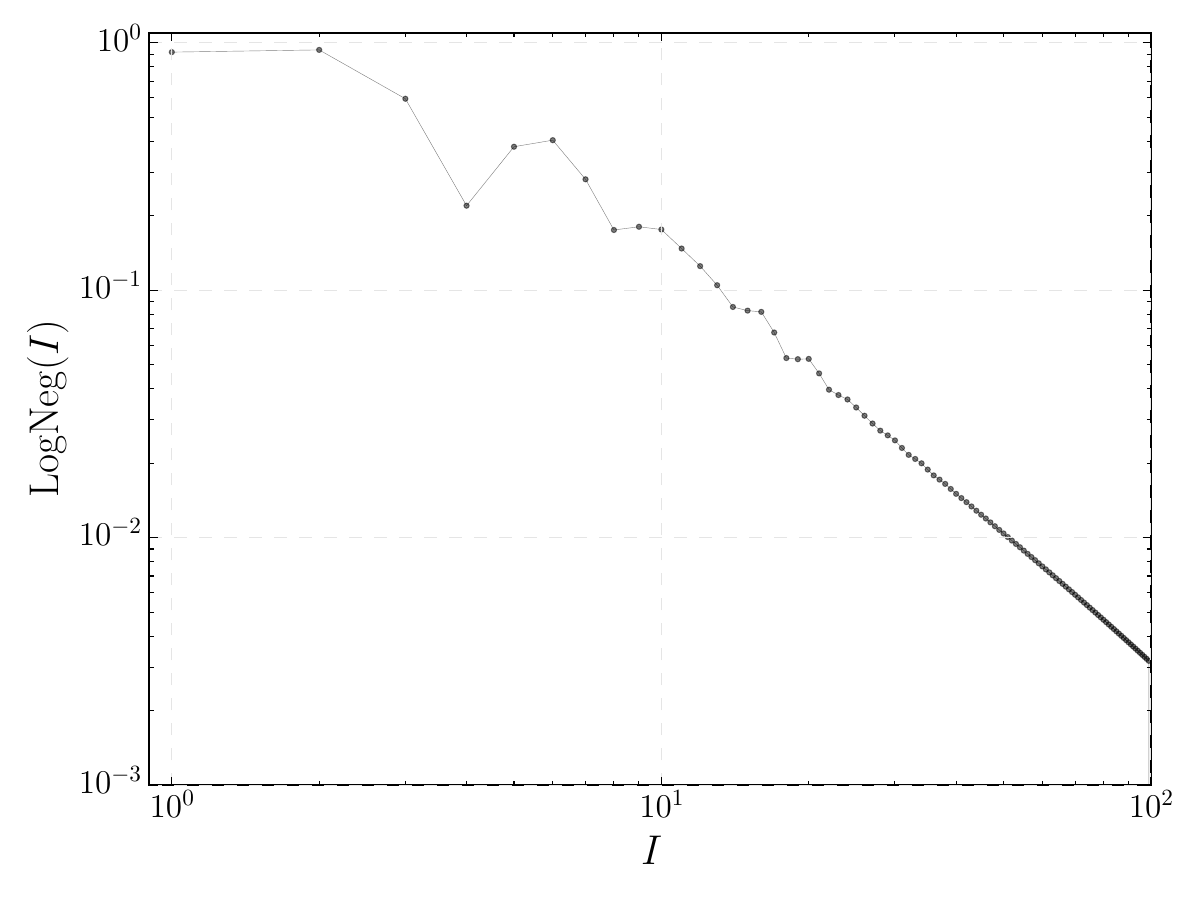}
}
\caption{\justifying Logarithmic negativity: this plot corresponds to the trajectories given by Eq. \eqref{eq:Traj-1plt}, with $\epsilon=0.375$ and $\kappa=33.3$. It shows the logarithmic negativity between a given Hawking mode $I$ and its partner in the limit $N\to\infty$ via Richardson extrapolation  out of simulations with total numbers of modes given by $N=256,\,512,\,1024$.}
\label{fig:1plt-logneg}
 \end{figure}

\jao{\it Quantum entanglement of vacuum:} Now, let us analyze quantum entanglement in the system. We have computed the logarithmic negativity of each Hawking mode $J=1,\dots,$ with its partner for an initial vacuum state. Namely, we set the subsystem $A$ to be a Hawking mode and subsystem $B$ its partner. Along with this calculation, we have also computed the logarithmic negativity for a configuration where we choose the same mode as subsystem $A$ and all other $N-1$ modes as subsystem $B$. We will denote it as $1\times(N-1)$ logarithmic negativity. For this purpose, we have developed the Python-based library \cite{montes2025}. In both cases we obtain identical results, as expected. In summary, a given mode as subsystem $A$ will have the same logarithmic negativity if we choose as subsystem $B$ the rest of the $(N-1)$ modes or, equivalently, its partner. By definition, the latter is the mode that purifies the Hawking mode once we trace out all other modes.  \jao{In Fig.~\ref{fig:1plt-logneg} we show the logarithmic negativity (quantum entanglement) between Hawking modes $I=1,\dots,$ and their partners for an initial vacuum state in the limit $N\to\infty$, extracted from simulations with $N=256,\,512,\,1024$ total modes using a Richardson extrapolation.} In summary, quantum entanglement oscillates with the same peaks and valleys as the ones in $|\beta_{\perp,J}|^2$ and $|\delta_{\parallel,J}|^2$, and it decreases toward the ultraviolet modes \jao{approximately as $(\omega_I)^{-2}$}, indicating that it is more difficult to entangle pairs of particles of higher energy. \jao{We have also computed the difference 
\begin{equation}\label{eq:Delta}
    \Delta_N(I) = |{\rm LogNeg}_{{}_{HP}}(I)-{\rm LogNeg}_{{}_{1\times(N-1)}}(I)| 
\end{equation}
between both calculations ($1\times(N-1)$ logarithmic negativity vs. logarithmic negativity of the Hawking-partner modes) for each Hawking mode $I$. We have checked that the two computations yield the same results up to numerical errors, for $N=256,\,512,\,1024$. Concretely, we have found $\Delta_{N=256}(I)\simeq 2\cdot 10^{-8} $, $\Delta_{N=512}(I)\simeq 4\cdot 10^{-8} $ and $\Delta_{N=1024}(I)\simeq 9\cdot 10^{-8} $ for all Hawking modes $I$.} However, there is an important difference: computational efficiency. We have realized that extracting logarithmic negativity following the HSU formula is considerably more efficient numerically than computing it directly from a $(1\times(N-1))$ bipartite state. Just to show an example, extracting logarithmic negativity from all bipartitions of the form $(1\times(N-1))$ for simulation of $N=1024$ can take many hours to several days. Following the HSU formula for all $out$ modes and computing the logarithmic negativity between all pairs of Hawking-partner modes can take just few minutes!

\begin{figure}[ht]
{\centering     
  \includegraphics[width = 0.48\textwidth]{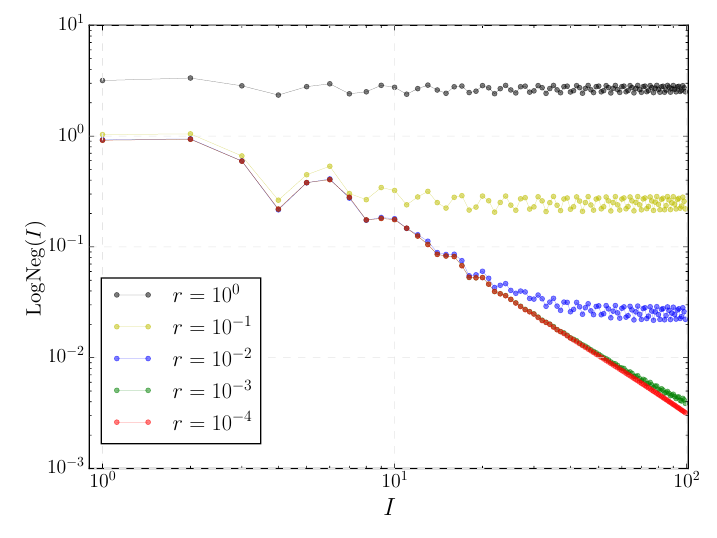}
}
\caption{\justifying $1\times(N-1)$ Logarithmic negativity: These plots correspond to the trajectories given by Eq. \eqref{eq:Traj-1plt}, with $\epsilon=0.375$ and $\kappa=33.3$. It shows the limit $N\to\infty$ via Richardson extrapolation, out of simulations with $N=256,\,512,\,1024$, for five initial one-mode squeezed states with squeezing intensities $r=10^{-4}$, $r=10^{-3}$, $r=10^{-2}$, $r=10^{-1}$, and $r=1$.}
\label{fig:1plt-1sq}
 \end{figure}

\jao{\it Quantum entanglement for squeezed states:} We have also considered initial states that are not the vacuum state. Concretely, we have studied initial states with some nonvanishing squeezing. We want to check if quantum entanglement can be stimulated  by this kind of trajectory of the boundary. Let us start with an initial state with all modes in a one-mode squeezed state (see Sec. \ref{sec:gauss-entangl} for their definition), with the same squeezing intensity. \jao{We carried out simulations for $N=256,\,512,\,1024$ and initial} states with all modes in a one-mode squeezed state. We consider the $1\times(N-1)$ entanglement entropy for three choices of initial squeezing intensity. In the asymptotic past, this $1\times(N-1)$ entanglement entropy is identically zero: each mode is not entangled with any other. In the asymptotic future, we see that the structure of entanglement is different from that in the vacuum case, specially for ultraviolet modes. \jao{In Fig. \ref{fig:1plt-1sq} we show the $1\times(N-1)$ logarithmic negativity for several values of the initial squeezing intensity, in the limit of $N\to \infty$ via Richardson extrapolation extracted from simulations with $N=256,\,512,\,1024$.} Quantum entanglement is nearly constant in this ultraviolet sector with some small oscillations. However, in the infrared sector small initial squeezing intensities (lower than $10^{-1}$) do not affect the quantum entanglement structure. It is qualitatively similar to the one of an initial vacuum state. In summary, for these infrared modes and relatively small squeezing intensities, we do not see considerable stimulation of entanglement. However, for ultraviolet modes and relatively large squeezing intensity, one-mode quantum entanglement reaches a plateau with a value proportional to initial squeezing intensity (with some superpose oscillations). Actually, if the squeezing intensity is higher than the maximum value of $1\times(N-1)$ squeezing entanglement for the vacuum state, quantum entanglement is always nearly flat (with some oscillations) and its value is given by the initial squeezing intensity. Therefore, we do see redistribution and some stimulation of quantum \jao{entanglement, clearly seen} when we increase the initial squeezing intensity.  

\begin{figure}[ht]
{\centering     
  \includegraphics[width = 0.48\textwidth]{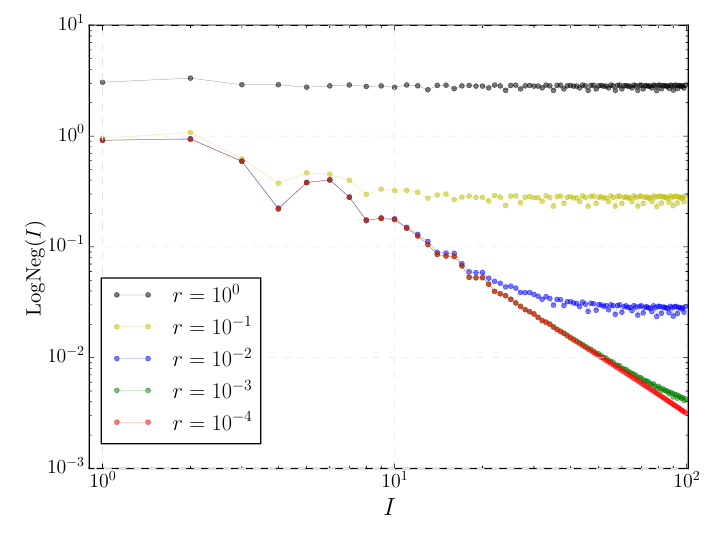}
}
\caption{\justifying $1\times(N-1)$ Logarithmic negativity: These plots correspond to the trajectories given by Eq. \eqref{eq:Traj-1plt}, with $\epsilon=0.375$ and $\kappa=33.3$. It shows the Logarithmic Negativity in the asymptotic future in the limit $N\to\infty$ via Richardson extrapolation (out of simulations with $N=256,\,512,\,1024$ total modes) for five initial two-mode squeezed states with squeezing intensities: $r=10^{-4}$, $r=10^{-3}$, $r=10^{-2}$, $r=10^{-1}$, and $r=1$.}
\label{fig:1plt-2sq}
 \end{figure}

For the sake of completeness we have also considered initial two-mode squeezed states with uniform squeezing intensity (see Sec. \ref{sec:gauss-entangl}). In this case, the initial $1\times(N-1)$ entanglement entropy will not be zero and will be independent of the partition of subsystem $A$ and $B$ (recalling that we always choose subsystem $A$ to be a given mode and $B$ the remaining $N-1$ modes). In the asymptotic future, $1\times(N-1)$ entanglement entropy shows the same qualitative structure as for initial one-mode squeezed states. The only difference we observe is that the amplitudes of the (small) oscillations are smaller in this case (see Fig. \ref{fig:1plt-2sq}). 

\begin{figure}[ht]
{\centering     
  \includegraphics[width = 0.48\textwidth]{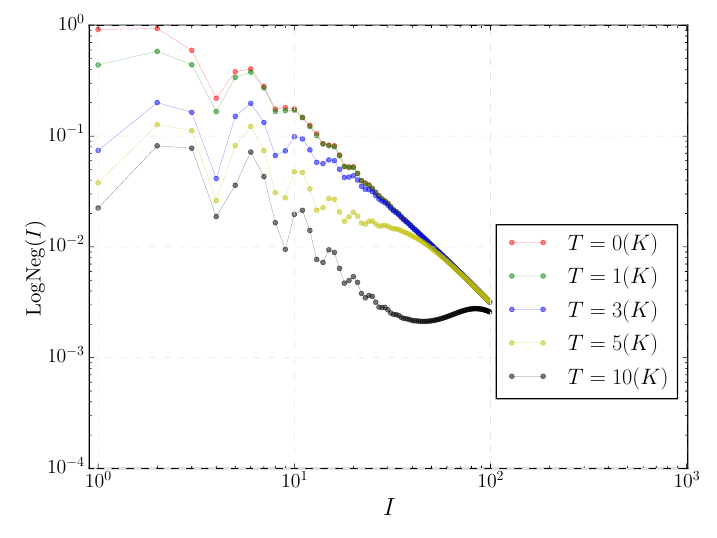}
}
\caption{\justifying $1\times(N-1)$ Logarithmic negativity: this plot corresponds to the trajectories given by Eq. \eqref{eq:Traj-1plt}, with $\epsilon=0.375$ and $\kappa=33.3$. It shows the Richardson extrapolation $N\to\infty$ (out of simulations with $N=256,\,512,\,1024$ total modes) for five initial thermal states with temperatures $T=0$ K (vacuum state), $T=1$ K, $T=3$ K, $T=5$ K, and $T=10$ K.}
\label{fig:1plt-thermal}
 \end{figure}

\jao{\it Quantum entanglement for thermal states:} We have seen that one of the advantages of the method based on the computation of the logarithmic negativity between the Hawking mode and its partner compared to the $1\times(N-1)$ logarithmic negativity is the computational efficiency. However, the first method can be used only for initial pure states. Hence, for initial mixed states, we can only extract the logarithmic negativity by means of the second method. This is the strategy we follow for the study of the entanglement entropy for initial thermal states. These mixed states will allow us to test the robustness of quantum entanglement against thermal noise, which will always be present in any experimental \jao{setting, as well as in astrophysical black holes due to the presence of cosmic microwave background}. 
\jao{In Fig. \ref{fig:1plt-thermal}, we show the $1\times(N-1)$ logarithmic negativity for five different initial thermal states with temperatures: $T=0$~K (vacuum state), $T=1$~K, $T=3$~K, $T=5$~K, and $T=10$~K. We show the Richardson extrapolation for $N\to\infty$ from simulations with total numbers of modes given by $N=256,\,512,\,1024$. Here, the $1\times(N-1)$ logarithmic negativity is robust for ultraviolet modes as we increase the temperature of the initial thermal state. However, the infrared modes show larger entanglement and more resilience against high thermal noise.  In all cases quantum entanglement shows oscillations. The local minima and maxima appear again at the minima and maxima of $|\beta_{\perp,J}|^2$ and $|\delta_{\parallel,J}|^2$. We have seen that, as we increase the temperature, the $1\times(N-1)$ entanglement entropy vanishes for some modes, and at sufficiently high temperatures, it completely vanishes. }

\jao{Actually}, we have found an interesting property: a critical temperature for quantum entanglement. Concretely, for each value of $N$ (the total number of modes), there is a critical temperature at which $1\times(N-1)$ logarithmic negativity vanishes. We have seen that this quantum entanglement exhibits a higher robustness to thermal noise as we increase the number of modes in the system. Concretely, we obtained $T_c(N=256)=27$ K, $T_c(N=512)=51$ K, and $T_c(N=1024)=96$ K. This almost linear behavior of the critical temperature with the number of modes appears to stem from an enhanced capacity of the system to store quantum entanglement as the number of degrees of freedom grows, particularly when the additional modes are associated with higher frequencies. Interestingly, if we define the critical temperature per mode ${\cal T}_c(N)=T_c(N)/N$, we have seen that in the limit $N\to\infty$ we obtain ${\cal T}_c=0.09145$ K. We have checked other configurations for the trajectories of the boundaries (where we change $\kappa$ and $\epsilon$) and we have obtained similar values for this critical temperature per mode in the limit $N\to\infty$, although the concrete numerical values change from one configuration to another. Actually, we provide a concrete example for a sharp trajectory below. 

Moreover, we have also seen that for all values of $N$ the partition that shows the strongest resilience against thermal noise corresponds in this case to mode $I=2$ as subsystem $A$.  Therefore, this partition is the only one where the last quantum entanglement survives until we reach the critical temperature. As we will see, this last property is not universal. On one hand, for small accelerations, the last quantum entanglement might not be stored in this concrete partition of the system, with other partitions showing stronger resistance to thermal noise. On the other hand, when accelerations of the boundaries are large, and hence their trajectories are sharper, it is for partitions with subsystem $A$ in the ultraviolet sector where quantum entanglement survives until one reaches the critical temperature. We will show a concrete example below.

\subsection{Large accelerations of the boundary}

Let us now discuss the configuration when the boundary follows a relatively large acceleration $\kappa$. In all cases we have found qualitatively similar results. Let us consider a concrete example given by $\epsilon=0.125$ and $\kappa=1200$. 

\begin{figure}[H]
{\centering     
  \includegraphics[width = 0.48\textwidth]{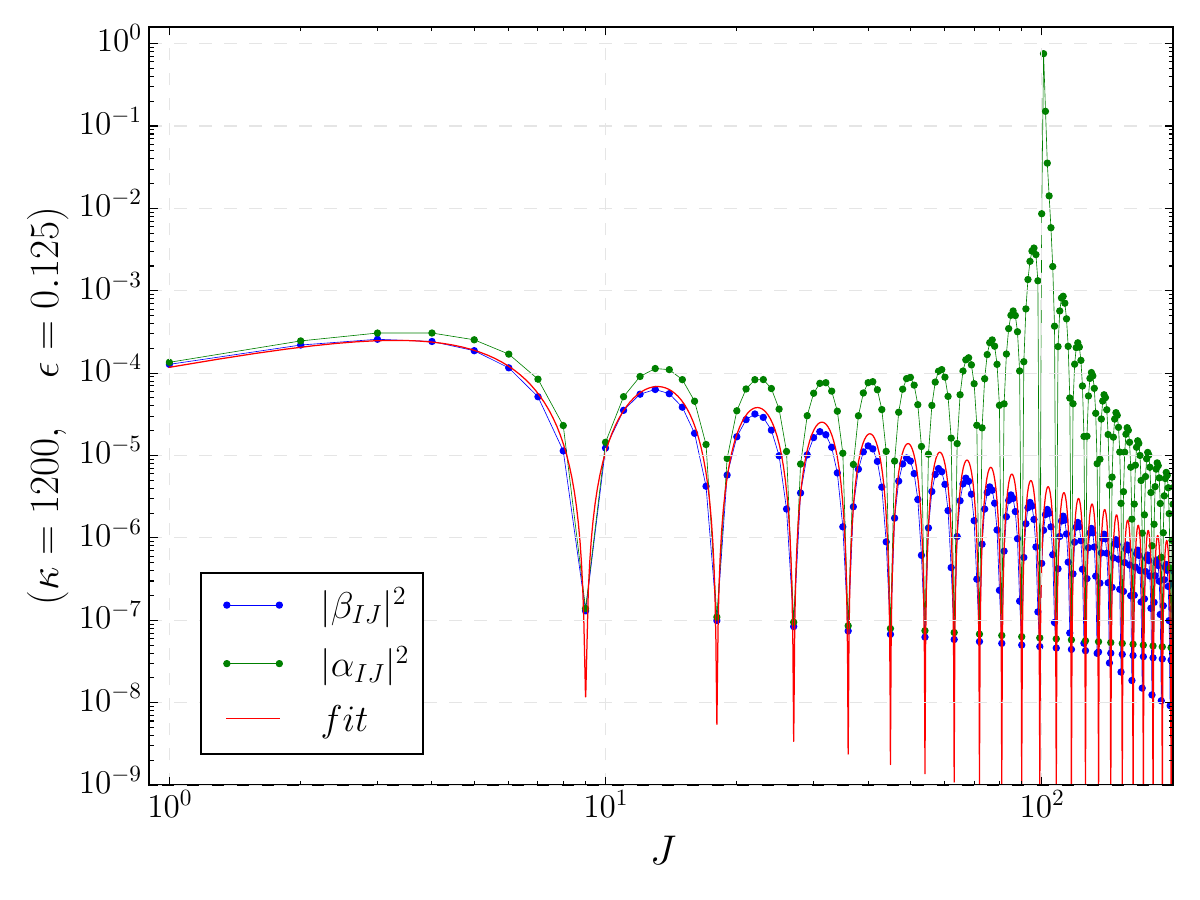}
  \includegraphics[width = 0.48\textwidth]{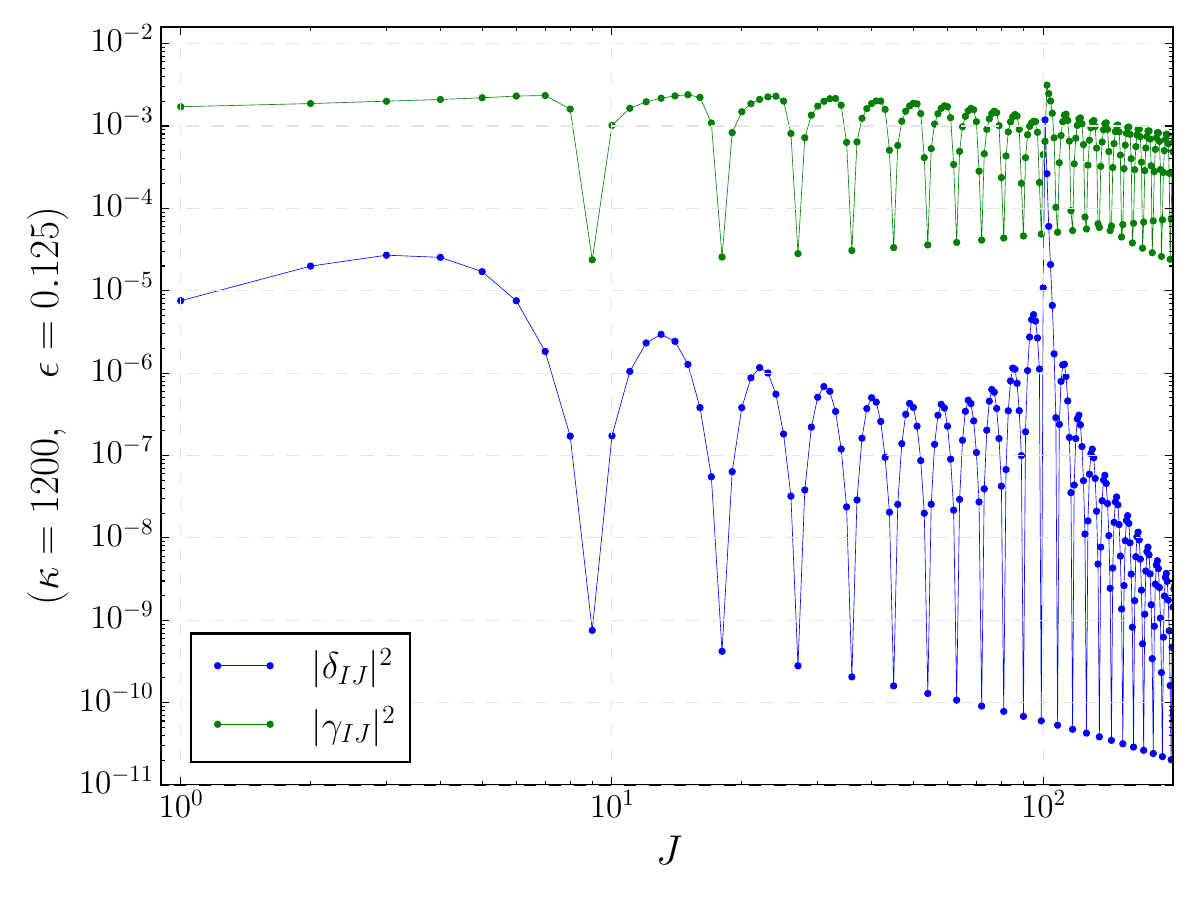}
}
\caption{\justifying Hawking and partner modes: these plots correspond to the trajectories given by Eq. \eqref{eq:Traj-1plt}, with $\epsilon=0.125$ and $\kappa=1200$. The upper panels show the (modulus squared of the) Bogoliubov coefficients of a Hawking mode (upper panel) and the ones of its partner (lower panel) for the $in$ frequency $I=90$ and $N\to\infty$ via Richardson extrapolation of the simulations with $N=256,\,512,\,1024$, and as functions of the $out$ frequencies in the interval $[1,100]$. }
\label{fig:1plt-HP-sharp}
\end{figure}

\jao{\it Bogoliubov coefficients:} In Fig. \ref{fig:1plt-HP-sharp} we show the Bogoliubov coefficients of a typical Hawking mode (upper panel) and its partner (lower panel). They are in the thermal frequency band \jao{(here we consider $I=90$ as an example)}, and we show the limit $N\to\infty$ after Richardson extrapolation of the $N=256,\,512,\,1024$ simulations. Together with the Hawking mode we also plot the fitting expression in Eq. \eqref{eq:betas-fit}. 

As shown, the Bogoliubov coefficients exhibit distinct behaviors for the Hawking and partner modes. For the Hawking mode, the $\beta$-coefficients match the fitting expression very well at low frequencies. At higher frequencies (not displayed in the figure), the Richardson extrapolation is not valid. We have seen that, on each set of simulations with $N=256,\,512,\,1024$, they follow a power-law decay and a final sharp suppression. In contrast, the $\alpha$-coefficients oscillate around a constant value (in logarithmic scale) at low frequencies, reach a peak near the corresponding $in$ mode (in this case, $J = 90$), and then drop rapidly. 
Eventually, as we said above, in the ultraviolet sector (not displayed in the figure), the Richardson extrapolation is not able to capture the physics. Here, we have seen that for each simulation with $N=256,\,512,\,1024$, they decay with the $out$ frequency according to a power law and a final sharp suppression.
The partner mode exhibits different trends. Its $\gamma$-coefficients, see Eq. \eqref{eq:partners-J}, oscillate around a nearly constant value (in log scale) until $out$ frequencies become of the order of the $in$ frequency of the Hawking mode, then they show a mild decay of around 2 orders of magnitude and a final sharp decay in the most ultraviolet region. The $\delta$-coefficients also oscillate in the infrared band, 
\begin{figure}[H]
{\centering     
\includegraphics[width= 0.48\textwidth]{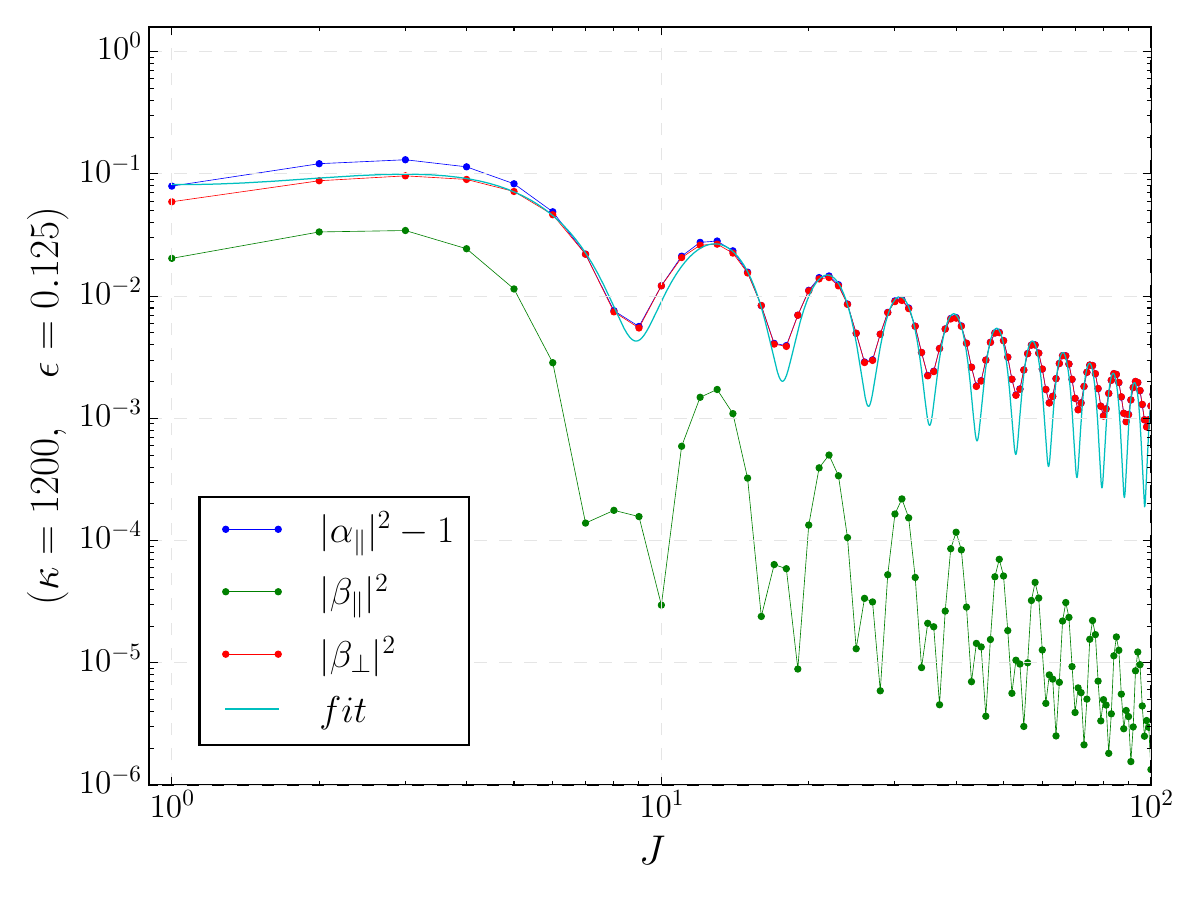}
\includegraphics[width = 0.48\textwidth]{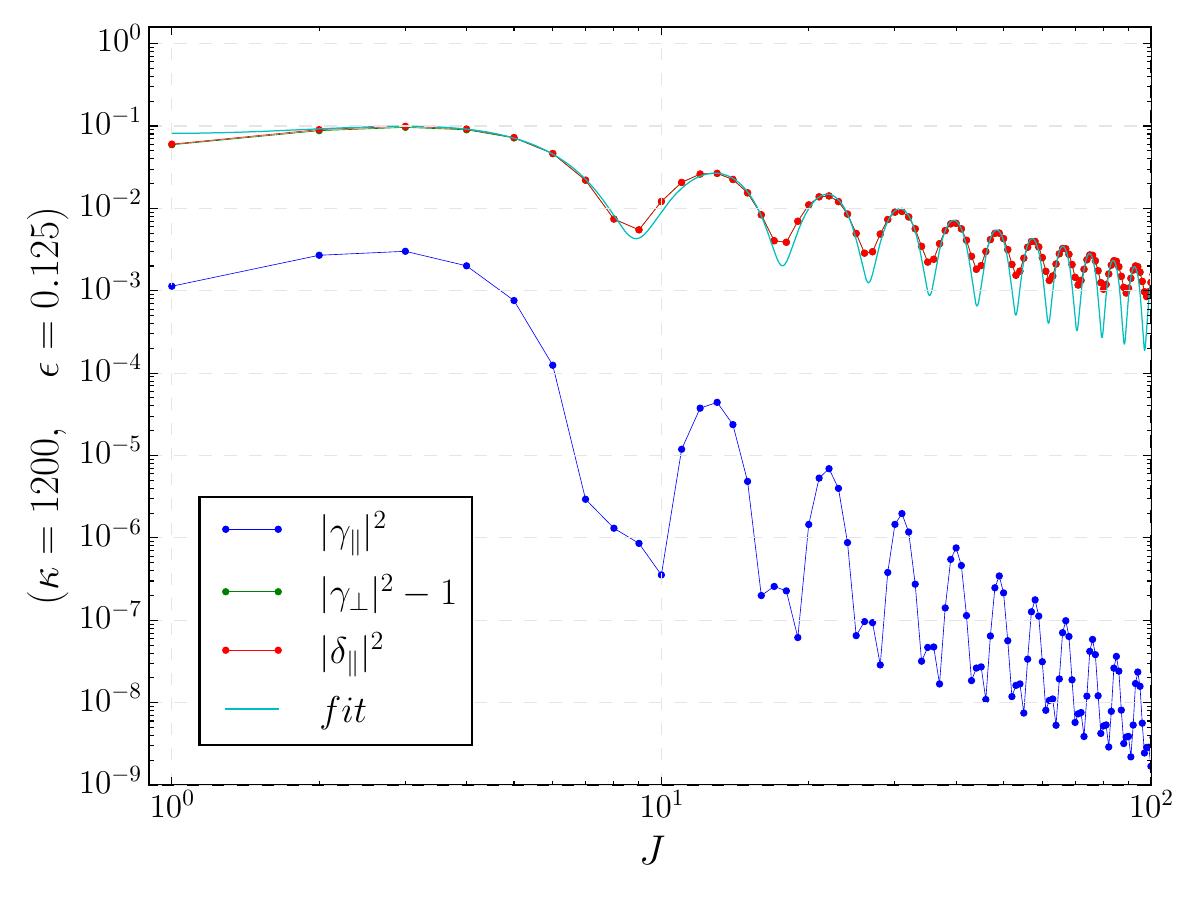}
}
\caption{\justifying Hawking and partner modes: these plots correspond to the trajectories given by Eq. \eqref{eq:Traj-1plt}, with $\epsilon=0.125$ and $\kappa=1200$. The upper panel shows the Bogoliubov coefficients in Eq. \eqref{eq:bh-mode} of all Hawking modes and the lower panel those of the partner modes in Eq. \eqref{eq:bp-mode}, as functions of the $out$ frequencies in the interval $[1,100]$ and in the limit $N\to\infty$ (Richardson extrapolation). Besides, we include in both cases the fitting expression in Eq.~\eqref{eq:betas-fit2}.}
\label{fig:1plt-HP-sharp-B}
\end{figure}
with a red tilt, peaking at $out$ frequencies near the $in$ Hawking mode (around $I \simeq 90$), before rapidly decaying at ultraviolet frequencies—at a rate faster than that of the $\beta$-coefficients of the Hawking mode.

In summary, $\gamma$-coefficients of the partner modes remain approximately constant across the spectrum, with their amplitudes differing by about 2 orders of magnitude between the infrared and ultraviolet frequencies. Meanwhile, the $\delta$-coefficients oscillate with a relatively constant amplitude in the infrared sector, show a peak near the Hawking mode frequency, and then drop sharply in the ultraviolet regime. This sharp suppression implies again that the partner modes contribute negligibly to high-energy particle content. Therefore, we can confidently conclude that, even for sharp trajectories, the purification of Hawking modes is predominantly a low-energy phenomenon \jao{(compared with the acceleration of the boundary or equivalently the surface gravity of the effective black hole).} Again, let us remember that we have fixed the Bogoliubov coefficients of partner modes under criterion {\bf B1} and we chose a family of symmetric trajectories of the boundaries. Other choices \jao{of trajectories} might yield different results. 

In Fig. \ref{fig:1plt-HP-sharp-B}, we also plot the Bogoliubov coefficients from Eqs. \eqref{eq:bh-mode} and \eqref{eq:bp-mode} as functions of the $out$ frequencies, assuming $\delta_{\perp} = 0$ throughout. We observe that, aside from the most infrared $out$ modes, the relation $|\alpha|^2 - 1 \simeq |\beta_\perp|^2$ holds, with $|\beta_\parallel|$ being negligible. Similarly, for the Hawking partners, we find $|\gamma_\perp|^2 - 1 \simeq |\delta_\parallel|^2$, with $|\gamma_\parallel|$ also negligible. These results reinforce again that, when these approximations are valid, each Hawking–partner pair effectively behaves as a two-mode squeezed state (see Sec. \ref{sec:gauss-entangl}). Furthermore, the coefficients $|\beta_\perp|^2$ and $|\delta_\parallel|^2$ show  agreement with the fitting expression of Eq.~\eqref{eq:betas-fit2}. 

We have also analyzed criterion {\bf B2} (i.e., $\boldsymbol{\gamma}\!\!\parallel\!\! \boldsymbol{\beta}$) and we have seen that only few Hawking modes fulfill it. Concretely, those modes around the local minima of $\beta_{\parallel}$ (see the upper panel of Fig. \ref{fig:1plt-HP-sharp-B}). Otherwise, only criterion {\bf B1} is valid. Since both options give the same results \jao{regarding quantum entanglement} whenever they are simultaneously applicable, we restrict here the study to criterion {\bf B1}. 

\begin{figure}[ht]
{\centering     
\includegraphics[width = 0.48\textwidth]{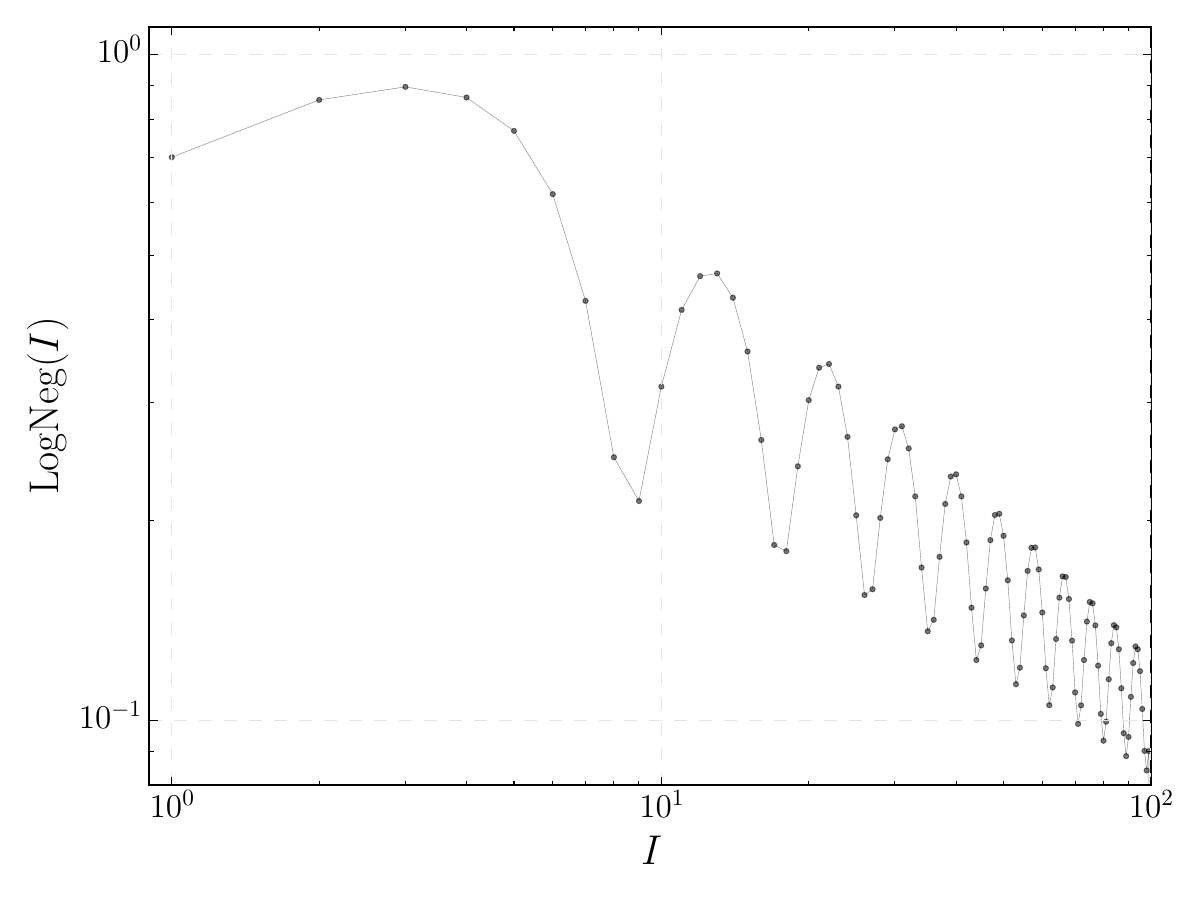}
}
\caption{\justifying Logarithmic Negativity: this plot corresponds to the trajectories described by Eq. \eqref{eq:Traj-1plt}, with parameters $\epsilon = 0.125$ and $\kappa = 1200$. It displays the logarithmic negativity between the Hawking mode $I=1,2,\ldots$ and its partner in the limit $N \to \infty$ after considering a Richardson extrapolation out of simulations with $N = 256, 512, 1024$.}
\label{fig:1plt-logneg-sharp}
 \end{figure}

\jao{\it Quantum entanglement of vacuum:} Moreover, we have computed the logarithmic negativity of each Hawking mode $I=1,\dots,$ with its partner for an initial vacuum state. We also compare the results of this calculation with the $1\times(N-1)$ logarithmic negativity with mode $I$ as subsystem $A$ and all other $N-1$ modes as subsystem $B$. In both cases we obtain, up to numerical errors, identical results. \jao{In Fig. \ref{fig:1plt-logneg-sharp}} we show the logarithmic negativity (quantum entanglement) between Hawking modes $I=1,\dots,$ and their partners for an initial vacuum state \jao{and only show the Richardson extrapolation in the limit $N\to\infty$ out of simulations with $N=256,\,512,\,1024$.} As we see, quantum entanglement shows oscillations whose maxima and minima coincide with those of $|\beta_\perp|^2$ and $|\delta_\parallel|^2$. In addition, it decreases toward the ultraviolet modes, indicating that the more energetic the modes are, the more difficult it is to entangle them. \jao{We have also computed $\Delta(I)$, defined in Eq. \eqref{eq:Delta}, between both calculations, i.e. logarithmic negativity between the Hawking mode and its partner vs. $1\times(N-1)$ logarithmic negativity of a Hawking mode and the rest of the system. We obtain $\Delta_{N=256}(I)\simeq 2\cdot 10^{-8} $, $\Delta_{N=512}(I)\simeq 3\cdot 10^{-8} $ and $\Delta_{N=1024}(I)\simeq 7\cdot 10^{-8} $ for all Hawking modes $I$.}

\begin{figure}[ht]
{\centering   
  \includegraphics[width = 0.48\textwidth]{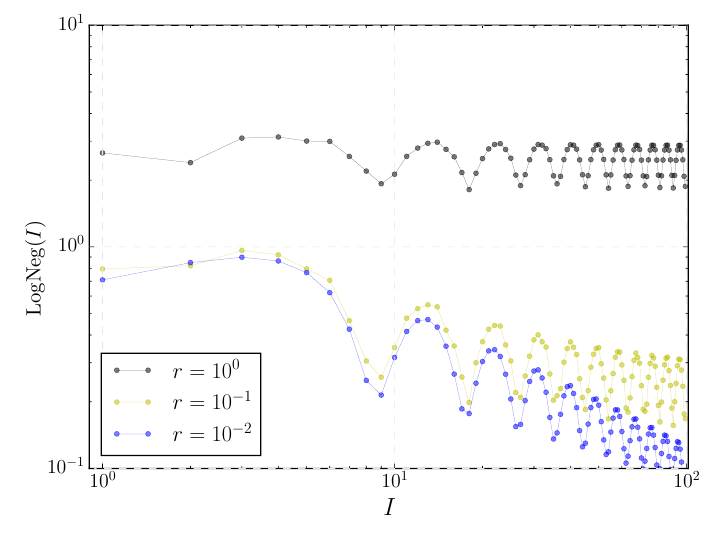}
}
\caption{\justifying $1\times(N-1)$ Logarithmic negativity: this plot corresponds to the trajectories given by Eq. \eqref{eq:Traj-1plt}, with $\epsilon=0.125$ and $\kappa=1200$. The dots  correspond to three different initial one-mode squeezed states with squeezing intensity $r=10^{-2}$, $r=10^{-1}$ and $r=1$. We show the limit $N\to\infty$ via Richardson extrapolation from simulations with $N=256,\,512,\,1024$. }
\label{fig:1plt-sharp-1sq}
 \end{figure}

\jao{\it Quantum entanglement for squeezed states:}  We have also considered squeezed initial states. Let us start with initial one-mode squeezed states (see Sec. \ref{sec:gauss-entangl} for their definition), with uniform squeezing intensity. In Fig. \ref{fig:1plt-sharp-1sq} we show the $1\times(N-1)$ logarithmic negativity for several values of the initial squeezing intensity. For the initial squeezing, we make three choices of intensity. Let us recall that, in the asymptotic past, this $1\times(N-1)$ entanglement entropy is identically zero since different modes are not initially entangled. In the asymptotic future, we see that the structure of entanglement is different from that of the vacuum case. For small initial squeezing intensities (lower than $10^{-1}$), quantum entanglement shows almost no dependence with the initial squeezing intensity at infrared frequencies, being qualitatively similar to the one of an initial vacuum state. However, for ultraviolet frequencies and relatively large squeezing intensity, one-mode quantum entanglement reaches a plateau with a value proportional to the initial squeezing intensity (with some superpose oscillations). Actually, if the squeezing intensity in the asymptotic past is higher than the maximum value of $1\times(N-1)$ squeezing entanglement for the vacuum state, the resulting quantum entanglement in the asymptotic future is always, up to some oscillations, nearly flat, and its value close but higher than the initial squeezing intensity. Therefore, we do see redistribution and some stimulation of quantum entanglement, since quantum correlations become stronger compared with the vacuum case, affecting mainly the ultraviolet sector. \jao{In all cases we show the limit $N\to \infty$ via Richardson extrapolation of $1\times(N-1)$ logarithmic  negativity out of simulations with $N=256,\,512,\,1024$}. 

\begin{figure}[ht]
{\centering     
  \includegraphics[width = 0.48\textwidth]{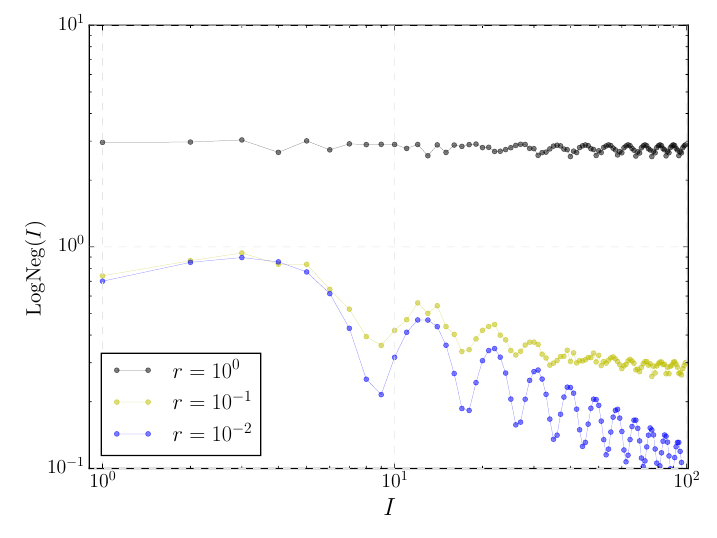}
}
\caption{\justifying $1\times(N-1)$ Logarithmic negativity: this plot corresponds to the trajectories given by Eq. \eqref{eq:Traj-1plt}, with $\epsilon=0.125$ and $\kappa=1200$. The dots correspond to three different initial two-mode squeezed states with squeezing intensity $r=10^{-2}$, $r=10^{-1}$ and $r=1$. We only show the limit $N\to\infty$ via Richardson extrapolation of simulations with total numbers of modes $N=256,\,512,\,1024$. }
\label{fig:1plt-sharp-2sq}
 \end{figure}

For the sake of completeness, we have also considered initial two-mode squeezed states with uniform squeezing intensity. The initial $1\times(N-1)$ logarithmic  negativity will not be zero now and it will be independent of subsystem $A$ provided subsystem $B$ is composed by $(N-1)$ modes. In the asymptotic future, we plot the $1\times(N-1)$ logarithmic  negativity in Fig. \ref{fig:1plt-sharp-2sq}. It shows the same qualitative structure as for initial one-mode squeezed states. The only difference we observe is that the amplitude of the oscillations is smaller in this case. 

\begin{figure}[ht]
{\centering     
  \includegraphics[width = 0.48\textwidth]{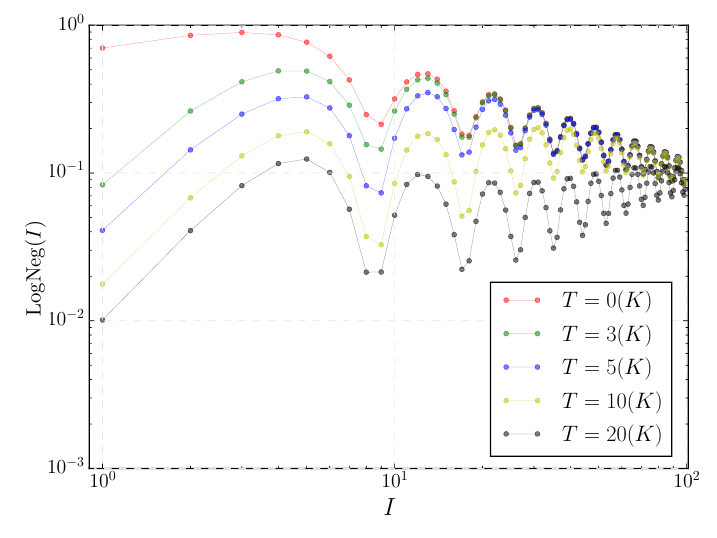}
}
\caption{\justifying $1\times(N-1)$ Logarithmic negativity: this plot corresponds to the trajectories given by Eq. \eqref{eq:Traj-1plt}, with $\epsilon=0.125$ and $\kappa=1200$. It shows the limit $N\to\infty$ via Richardson extrapolation of simulations with $N=256,\,512,\,1024$, for five initial thermal states with temperatures $T=0$ K (vacuum state), $T=3$ K, $T=5$ K, $T=10$ K, and $T=20$ K. }
\label{fig:1plt-sharp-thermal}
 \end{figure}

\jao{\it Quantum entanglement for thermal states:} In addition to the results obtained for initial pure states, in Fig. \ref{fig:1plt-sharp-thermal} we plot the $1\times(N-1)$ entanglement entropy for five different initial thermal states with temperatures: $T=0$ K (vacuum state), $T=3$ K, $T=5$ K, $T=10$ K and $T=20$ K. In all cases we show the extrapolation $N\to\infty$ extracted from simulations with total number of modes given by $N=256,\,512,\,1024$, via Richardson extrapolation. As we see, increasing the temperature results in a lower $1\times(N-1)$ entanglement entropy, especially in the infrared sector. We have also checked that, at sufficiently high temperatures, this quantum entanglement completely vanishes. Again, the corresponding critical temperature depends on the total number of modes. We have found that this quantum entanglement is larger as we increase the value of $\kappa$ and it is also stronger against thermal noise as we increase the number of modes in the system. Concretely, we obtained $T_c(N=256)=36$ K, $T_c(N=512)=68$ K, and $T_c(N=1024)=128$ K. Hence, it is a monotonic growing function with $N$. Interestingly, following the definition of the critical temperature per mode ${\cal T}_c(N)=T_c(N)/N$, in the limit $N\to\infty$ we obtain ${\cal T}_c=0.1219$ K. We have checked other configurations for the trajectories of the boundaries (where we change $\kappa$ and $\epsilon$) for this regime of sharp trajectories, and we obtain similar values for this critical temperature per mode in the limit $N\to\infty$. The concrete numerical values depend on the particular choices of $\kappa$ and $\epsilon$.

In all cases, quantum entanglement shows oscillations. Again, the local minima and maxima appear at the minima and maxima of $|\beta_\perp|^2$ and $|\delta_\parallel|^2$. This indicates that frequency channels with a more efficient particle production also produce particle pairs with stronger quantum correlations.
Unlike in the previous case (with small values of $\kappa$), we have not found a concrete partition where $1\times(N-1)$ logarithmic negativity survives until we reach the critical temperature. Different values of $N$ have the last quantum entanglement stored in different partitions. Concretely, for this example, for $N=256$ and $N=512$, the last quantum entanglement is stored in the ultraviolet sector, while for $N=1024$ this happens for infrared modes. 

\section{Discussion and conclusions}\label{sec:concl}

In this work we investigate quantum entanglement generated in a one-dimensional cavity where one boundary undergoes a prescribed acceleration, mimicking aspects of Hawking radiation in analogue gravity systems. \jao{Previous studies \cite{GarciaMartin-Caro:2023jjq,GarciaMartin-Caro:2025xgr} were primarily focused on the analysis of Bogoliubov coefficients, rather than quantum entanglement.} Our analysis is based on Gaussian state theory. These states are fully characterized by their mean vector and covariance matrix, making the analysis computationally straightforward.
We have developed (open source) numerical tools for this purpose. For the time evolution of the system, we decompose the scalar field in modes. We then introduce a cutoff in the total number of modes and study their evolution (integrating a finite number of ordinary differential equations). Finally, we take the limit of infinite modes using Richardson extrapolation.

For an initial vacuum state, the motion of the boundary induces a change in the field's vacuum, leading to particle creation, with a nearly thermal spectrum in the infrared frequency band, resembling the Hawking effect up to a greybody factor that appears due to the finiteness of the motion of the boundary. The Bogoliubov coefficients, particularly $\beta_{IJ}$, describe this particle production and mode mixing, and \jao{together $\alpha_{IJ}$, } the transformation between the $in$ and $out$ states. Concretely, in both small and large acceleration regimes \jao{of the boundary}, the $\beta$-coefficients of the Hawking modes at low frequencies align well with a  Fulling-Davies spectrum with a greybody factor. 

We have also computed the partner modes of each Hawking mode, which are constructed following the Hotta-Sch\"utzhold-Unruh formula \jao{within criterion {\bf B1}}. Their $\delta$-coefficients oscillate, peak around the Hawking mode's frequency, and then sharply suppress in the ultraviolet sector. This indicates that the {\it purification of Hawking modes is primarily a low-energy process}, with almost no high-energetic particles associated with either Hawking or partner modes. One must recall that this statement is only valid for the particular trajectories of the boundaries considered in this \jao{manuscript, which mimic the formation of a black hole that transforms into a white hole with the same surface gravity which totally evaporates}. For instance, a sudden stop with a sharp transient will likely produce high-energy particles. \jao{This would correspond to the formation of a black hole that suddenly evaporates. The} analysis also explicitly revealed that, in cases where certain approximations hold, each Hawking-partner pair effectively behaves as a two-mode squeezed state. This leads to the conclusion that an expanding cavity acts as a multimode squeezing device. \jao{We must note that the Hotta-Sch\"utzhold-Unruh formula is not free of ambiguities, where we fix them following a physically motivated criterion. Therefore, some of the above statements might be sensitive to this freedom. }

\jao{We have also checked that, for initial pure states,} the logarithmic negativity calculated between a Hawking mode and its partner is identical (up to numerical errors) to the $1 \times (N-1)$ logarithmic negativity where the same mode is taken as subsystem $A$ and all other modes as subsystem $B$. This, as expected, supports the idea that the partner mode precisely purifies the Hawking mode when all other degrees of freedom are traced out. In both acceleration regimes, quantum entanglement {\it decreases toward the ultraviolet modes}, suggesting that it is more challenging to entangle pairs of higher-energy particles. In all cases, it exhibits oscillations, with local minima and maxima appearing at the minima and maxima of particle production. This suggests that frequency channels with more efficient particle production also generate particle pairs with stronger quantum \jao{correlations, which are largest for infrared modes.} In fact, all these properties indicate that the ambiguities of the construction \cite{hotta2015Partner} will not affect the main conclusions reached \jao{here: the related computations involve symplectic invariants that are not sensitive to the freedom in the construction of the Hawking partners.} Actually, the results found in Ref. \cite{GarciaMartin-Caro:2025xgr} about the robustness of the thermal spectrum for infrared frequencies with respect to \jao{more general} trajectories of the boundary \jao{than the ones discussed here} suggest that a detailed analysis of the structure of quantum entanglement \jao{in these settings} seems necessary. Our preliminary calculations indicate deviations mainly in the ultraviolet sector. They will be discussed in detail in a future publication. 

Our study also explored the impact of nonvacuum initial states, specifically one-mode and two-mode squeezed states, and thermal states, on quantum entanglement. For initial one-mode squeezed states and relatively small initial squeezing intensities, quantum entanglement shows little influence from the initial squeezing and retains a structure qualitatively similar to the vacuum state at infrared frequencies. However, for relatively large squeezing intensities, {\it quantum entanglement is redistributed and stimulated}, reaching a plateau proportional to the initial squeezing intensity, particularly affecting the ultraviolet modes, and with the proportionality constant larger than unit. Initial two-mode squeezed states also lead to entanglement redistribution and stimulation with a qualitatively similar structure to the one of one-mode squeezed states, though with smaller oscillation amplitudes. In all cases with initial squeezing, quantum entanglement exhibits oscillations, as in the vacuum case.  Finally, thermal initial states can mimic how thermal noise affects entanglement, a crucial question in analogue gravity scenarios \jao{and realistic astrophysical black holes that are immersed in a thermal bath due to the cosmic microwave background}. As the initial temperature of the thermal state increases, the $1 \times (N-1)$ entanglement entropy decreases, especially in the infrared modes, with quantum entanglement being very robust in the ultraviolet sector, and eventually vanishing completely at sufficiently high but finite temperatures. We have found the corresponding  critical temperature \jao{at which quantum entanglement is completely degraded}. It increases linearly with the total number of modes, suggesting enhanced resilience to thermal noise with more degrees of freedom as a consequence of the multimode entanglement structure of the final state. This behavior allowed us to introduce the critical temperature per mode, \jao{which gives a finite result in the $N \to \infty$ limit.} The values of this quantity depend on the specific boundary trajectory parameters, but with very similar magnitudes, indicating that the critical temperature per mode has a weak dependence on the acceleration of the boundaries. Besides, we have also seen that the strongest resilience to thermal noise was found for modes in the infrared sector if trajectories of the boundaries involve small accelerations. For large accelerations, the last remaining entanglement can be stored in different partitions (infrared or ultraviolet) depending on $N$. In all cases with thermal noise, quantum entanglement exhibits oscillations, with local minima and maxima appearing at the minima and maxima of particle production, as in the vacuum case. 

Finally, we want to stress that the Hotta-Sch\"utzhold-Unruh formula offers significant computational efficiency regarding the evaluation of the logarithmic negativity for pure states, drastically reducing computation times compared to calculating the $1 \times (N-1)$ logarithmic negativity. Furthermore, the commutators between Hawking and partner annihilation operators for different modes generally do not vanish, indicating a nontrivial multimode entanglement structure. Here, a basis of Hawking-partner modes in its present form is not suitable to fully describe the entire system. 

We finish this section with a short discussion about the potential consequences regarding the experimental verification of our findings. Our theoretical setting is the same as that used for the study of the dynamical Casimir effect. Its experimental observation has been studied in Refs.~\cite{Wilson_2011,Johansson:2009zz,Johansson:2010vqd,Doukas:2014bja,Pasi2013}. Quantum entanglement in this framework has been recently analyzed in Ref. \cite{Romualdo:2019eur}. The experimental setting has under control thermal noise up to temperatures around $T=0.025$ K. Besides, one should expect an ultraviolet cutoff in frequencies of the order of the plasma frequency of the SQUIDs, which in these cases is around $\omega_p=37.3$ GHz. For transmission lines with propagation speed $v\simeq 10^8$ m/s, as long as $L_0=0.1$ m and changes in cavity size of the order of $dL=0.025$ m, the maximum acceleration of the boundary will be $\kappa_{max} = \omega_p^2dL\simeq 3.5\cdot 10^{19}$ m/s${}^2$, and hence a maximum Hawking temperature $T_{max} \simeq 0.4$ K. Now, taking into account that the fundamental frequency in the cavity is $\omega_0\simeq 3.1\cdot 10^{19}$ GHz and the maximum frequency is $\omega_{max}=\omega_p$, the total number of modes in the cavity will be $N\simeq 10$. If we assume, based on our numerical results, that the critical temperature per mode (at which quantum entanglement completely disappears) is around ${\cal T}_c \simeq 0.1$ K, we find that the critical temperature for the experimental setting with the above parameters will be $T_c(N=10)=1$ K, assuming ${\cal T}_c\simeq T_c(N)/N$. In fact, we have performed a numerical simulation with $N=10$, $\epsilon = 0.375$ and $\kappa=33$, and we have obtained a critical temperature of $1.5$ K. This critical temperature is of the order (but above) of the thermal noise in the experimental settings. Hence, as was already mentioned in Ref. \cite{GarciaMartin-Caro:2023jjq}, it is possible not only to observe a thermal spectrum with the temperature given by the acceleration of the boundary but also to extract quantum entanglement, at least in theory, since it will not be completely disrupted by current experimental thermal noise.  

\acknowledgments

The authors would like to thank Iv\'an Agullo, \jao{Alberto Garc\'ia Mart\'in-Caro, and Jos\'e M. S\'anchez Vel\'azquez} for helpful discussions. Financial support is provided by the Spanish Government through the projects PID2020-119632GB-I00, and PID2019-105943GB-I00 (with FEDER contribution), and PID2022-140831NB-I00 funded by MI-CIU/AEI/10.13039/501100011033 and FEDER, UE. 

\appendix

\section{Some properties on the HSU algorithm}\label{app:HSU-alg}

This algorithm is very useful for analyzing some properties of single-by-single Hawking-partner pairs. However, it also has important limitations that we will discuss in this Appendix. Let us restrict the analysis to a given Hawking mode and its partner (for instance by tracing out all other degrees of freedom). The corresponding (restricted) mode field (and momenta) operator of a Hawking mode will have the form
\begin{equation}
    \hat{\bf U}_{H}(t)=\hat b_H {\bf w}_{H}(t)+\hat b_H^\dagger \bar{\bf w}_{H}(t).
\end{equation}
Now, taking into account Eq. \eqref{eq:bh-aI} and 
\begin{equation}
{\bf w}_{H}(t) = \sum_{I=1}^{N} \bar\alpha_{I}\;{}^{(in)}{\bf u}^{(I)}(t)-\beta_{I}\;{}^{(in)}\bar {\bf u}^{(I)}(t),
\end{equation}
we arrive at
\begin{equation}
    \hat{\bf U}_{H}(t)= \sum_I^{N} \hat a_I {\bf v}^{(I)}_{H}(t)+\hat a_I^\dagger \bar{\bf v}^{(I)}_{H}(t),
\end{equation}
such that 
\begin{equation}
{\bf v}^{(I)}_{H}(t) = \sum_{I'}^{N}\alpha^{H}_{II'}{}^{(in)}{\bf u}^{(I')}(t)+\beta^{H}_{II'}{}^{(in)}\bar{\bf u}^{(I')}(t),
\end{equation}
with
\begin{equation}
    \alpha^{H}_{II'}= \alpha_I\bar\alpha_{I'}-\beta_I\bar\beta_{I'},\quad \beta^{H}_{II'}=-\alpha_I\beta_{I'}+\beta_I\alpha_{I'}.
\end{equation}
We can now compute the norm of ${\bf v}^{(I)}_{H}(t)$. One can see that 
\begin{equation}
\langle {\bf v}^{(I)}_{H}(t),{\bf v}^{(I')}_{H}(t)\rangle = \alpha_I\bar\alpha_{I'}-\beta_I\bar\beta_{I'}.
\end{equation}
In general, the right-hand side of this equation is not equal to \jao{Kronecker delta}. We have checked that in general for $I=I'$ and most of the Hawking modes, it is greater than \jao{zero}. Therefore, ${\bf v}^{(I)}_{H}(t)$ are positive frequency solutions, but they do not provide a basis of solutions. 

By the same arguments, the partner mode has mode field (and momenta) operator
\begin{equation}
    \hat{\bf U}_{P}(t)=\hat b_P {\bf w}_{P}(t)+\hat b_P^\dagger \bar{\bf w}_{P}(t),
\end{equation}
and by Eq. \eqref{eq:bp-aI} and 
\begin{equation}
{\bf w}_{P}(t) = \sum_{I=1}^{N} \bar\gamma_{I}\;{}^{(in)}{\bf u}^{(I)}(t)-\delta_{I}\;{}^{(in)}\bar {\bf u}^{(I)}(t),
\end{equation}
we arrive at
\begin{equation}
    \hat{\bf U}_{P}(t)= \sum_I^{N} \hat a_I {\bf v}^{(I)}_{P}(t)+\hat a_I^\dagger \bar{\bf v}^{(I)}_{P}(t),
\end{equation}
where
\begin{equation}
{\bf v}^{(I)}_{P}(t) = \sum_{I'}^{N}\gamma^{P}_{II'}{}^{(in)}{\bf u}^{(I')}(t)+\delta^{P}_{II'}{}^{(in)}\bar{\bf u}^{(I')}(t),
\end{equation}
with
\begin{equation}
    \gamma^{P}_{II'}= \gamma_I\bar\gamma_{I'}-\delta_I\bar\delta_{I'},\quad \delta^{P}_{II'}=-\gamma_I\delta_{I'}+\delta_I\gamma_{I'}.
\end{equation}
If we compute the inner product of two of these solutions, we obtain
\begin{equation}
\langle {\bf v}^{(I)}_{P}(t),{\bf v}^{(I')}_{P}(t)\rangle = \gamma_I\bar\gamma_{I'}-\delta_I\bar\delta_{I'}.
\end{equation}
The right-hand side is not a Kronecker delta in general. Therefore, ${\bf v}^{(I)}_{P}(t)$ do not provide a basis of solutions, although we have checked that for all partner modes they have positive norm for all $I=I'$ and all the partner modes in our numerical simulations. 

One should expect that, after truncating the system to a Hawking-partner pair, one is also losing an important part of the information of the system contained in those truncated pairs.

\begin{figure}[H]
{\centering     
  \includegraphics[width = 0.48\textwidth]{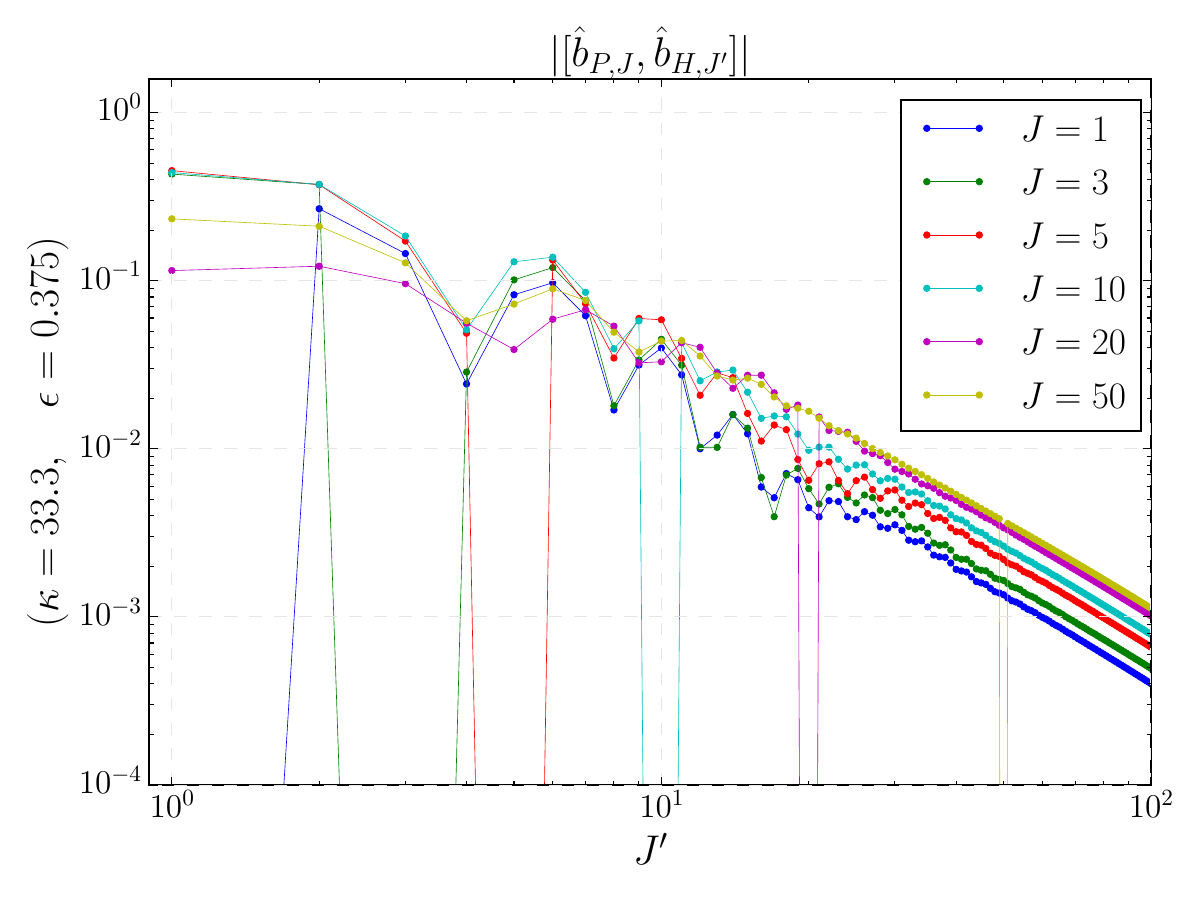}
  \includegraphics[width = 0.48\textwidth]{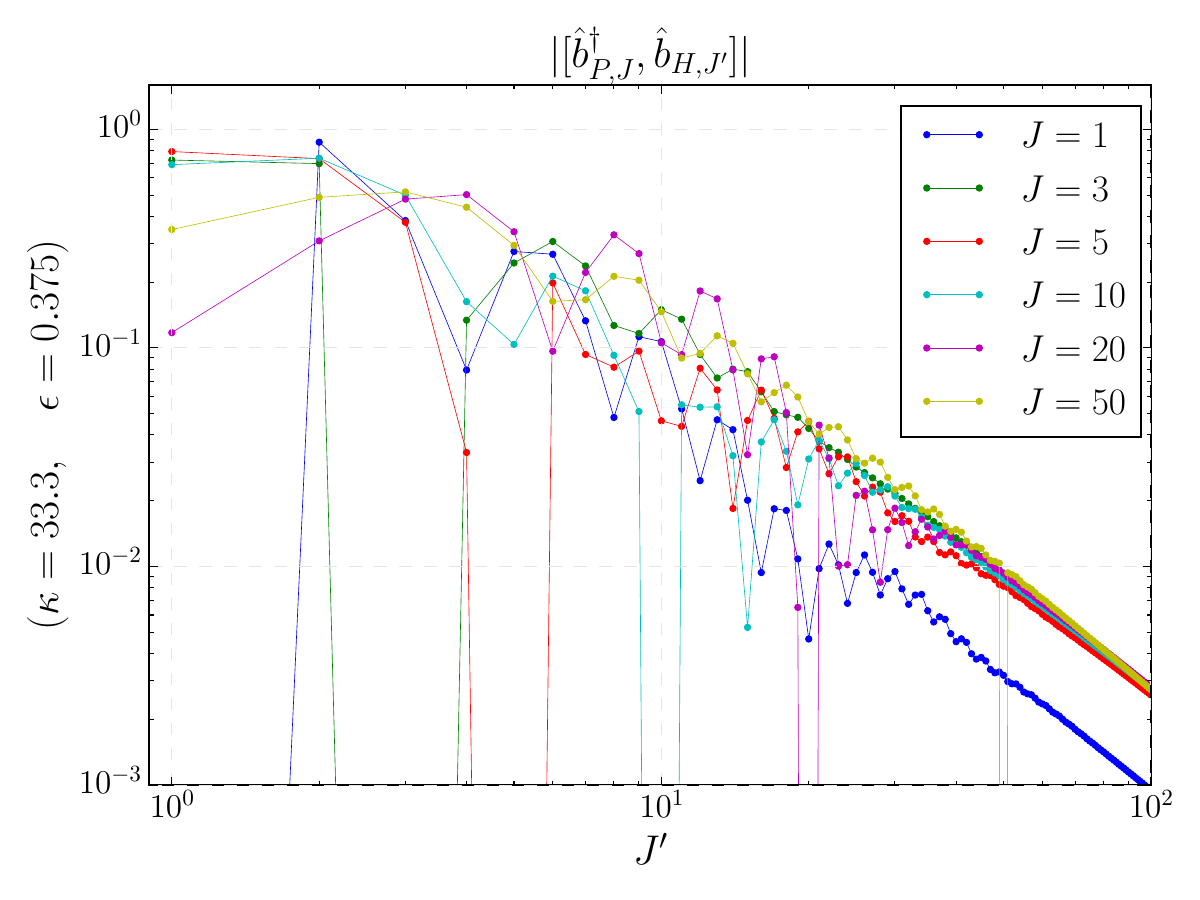}
}
\caption{\justifying Commutators of Hawking modes and partners: these plots correspond to the trajectories given by Eq. \eqref{eq:Traj-1plt}, with $\epsilon=0.375$ and $\kappa=33.3$. We show the (modulus of the) commutator in Eq. \eqref{eq:comm-bp-bh} (upper panel) and the one in Eq. \eqref{eq:comm-bpdagg-bh} (lower panel). We see that they obey the usual commutation relations if $J=J'$ (see the nearly vertical lines joining points with vanishingly small values). Otherwise, they do not vanish and are not negligible. This simulation corresponds to $N=1024$. Other choices of $N$ give qualitatively similar results. }
\label{fig:1plt-HP-comm}
\end{figure}

Besides, we have studied in some detail the commutators in Eq. \eqref{eq:comm-bpdagg-bh} as well as 

\begin{equation}\label{eq:comm-bpdagg-bh}
[\hat b^\dagger_{P,J}, \hat b_{H,J'}]=-\sum_I\bar\gamma_{IJ}\alpha_{IJ'}-\delta_{IJ}\bar\beta_{IJ'},    
\end{equation}
\begin{equation}\label{eq:comm-bp-bp}
[\hat b_{P,J}, \hat b_{P,J'}]=\sum_I-\gamma_{IJ}\bar\alpha_{IJ'}+\bar\delta_{IJ}\beta_{IJ'},    
\end{equation}
\begin{equation}\label{eq:comm-bpdagg-bp}
[\hat b^\dagger_{P,J}, \hat b_{P,J'}]=\sum_I-\bar\gamma_{IJ}\bar\gamma_{IJ'}+\delta_{IJ}\bar\delta_{IJ'}.    
\end{equation}

\begin{figure}[H]
{\centering     
  \includegraphics[width = 0.48\textwidth]{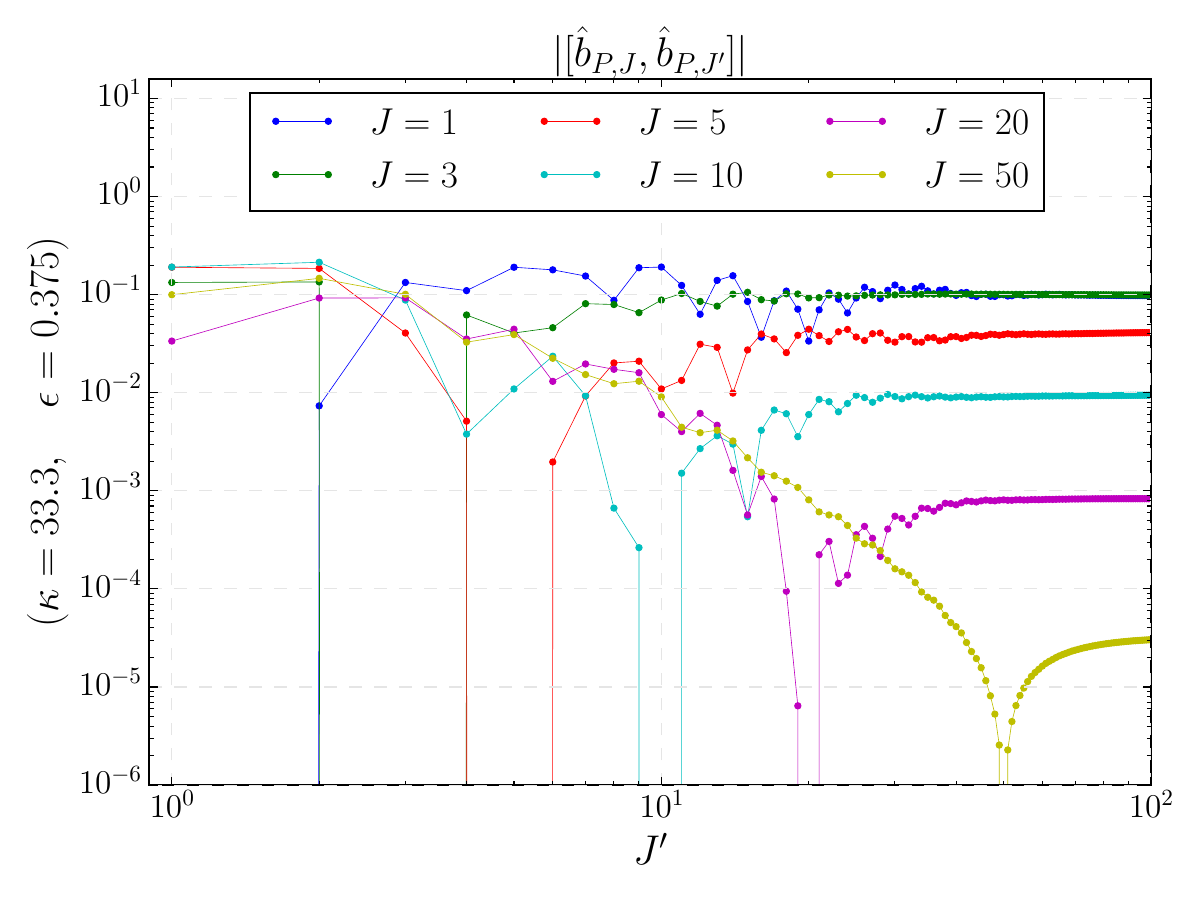}
  \includegraphics[width = 0.48\textwidth]{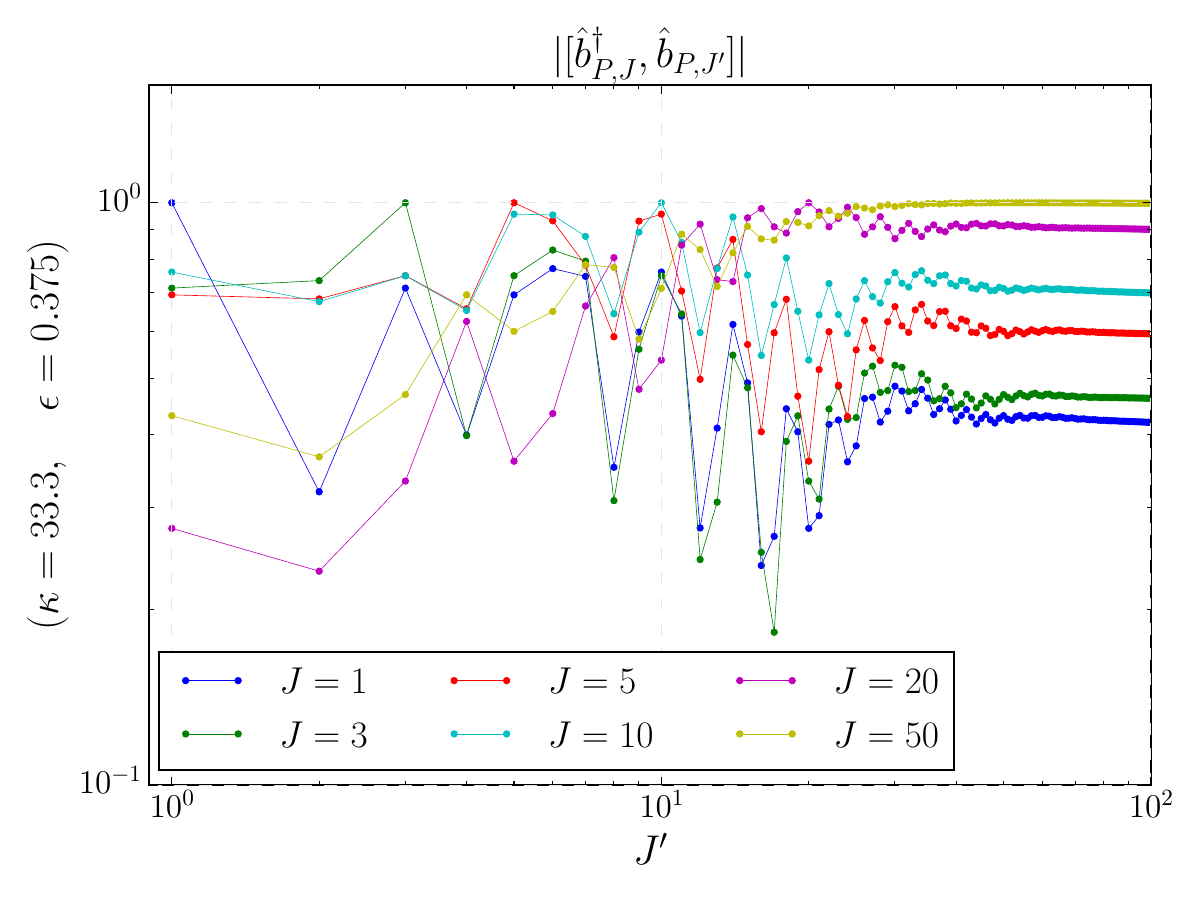}
}
\caption{\justifying Commutators of Hawking modes and partners: these plots correspond to the trajectories given by Eq. \eqref{eq:Traj-1plt}, with $\epsilon=0.375$ and $\kappa=33.3$. We show the (modulus of the) commutator in Eq. \eqref{eq:comm-bp-bp} (upper panel) and the one in Eq. \eqref{eq:comm-bpdagg-bp} (lower panel). We see that they obey the usual commutation relations if $J=J'$ (see the nearly vertical lines joining points with vanishingly small values). Otherwise, they do not vanish and are not negligible. This simulation corresponds to $N=1024$. Other choices of $N$ give qualitatively similar results. }
\label{fig:1plt-PP-comm}
\end{figure}
In all plots we show in this appendix, axes are in logarithmic scale. All points (the actual simulations) are joined by lines for illustrative purposes. In general we see that commutators are nonvanishing, except in some particular cases, and concretely, when $J = J'$.  They can be identified by the corresponding (almost) vertical lines in the plots. In Fig. \ref{fig:1plt-HP-comm} we show the above commutators with a concrete example for relatively small accelerations of the boundary. As we see,  these commutators do not vanish in general for $J\neq J'$, and in some cases they are of the order of the unit, as one can see in Fig. \ref{fig:1plt-PP-comm}, where we show the remaining commutators.

\begin{figure}[H]
{\centering     
  \includegraphics[width = 0.48\textwidth]{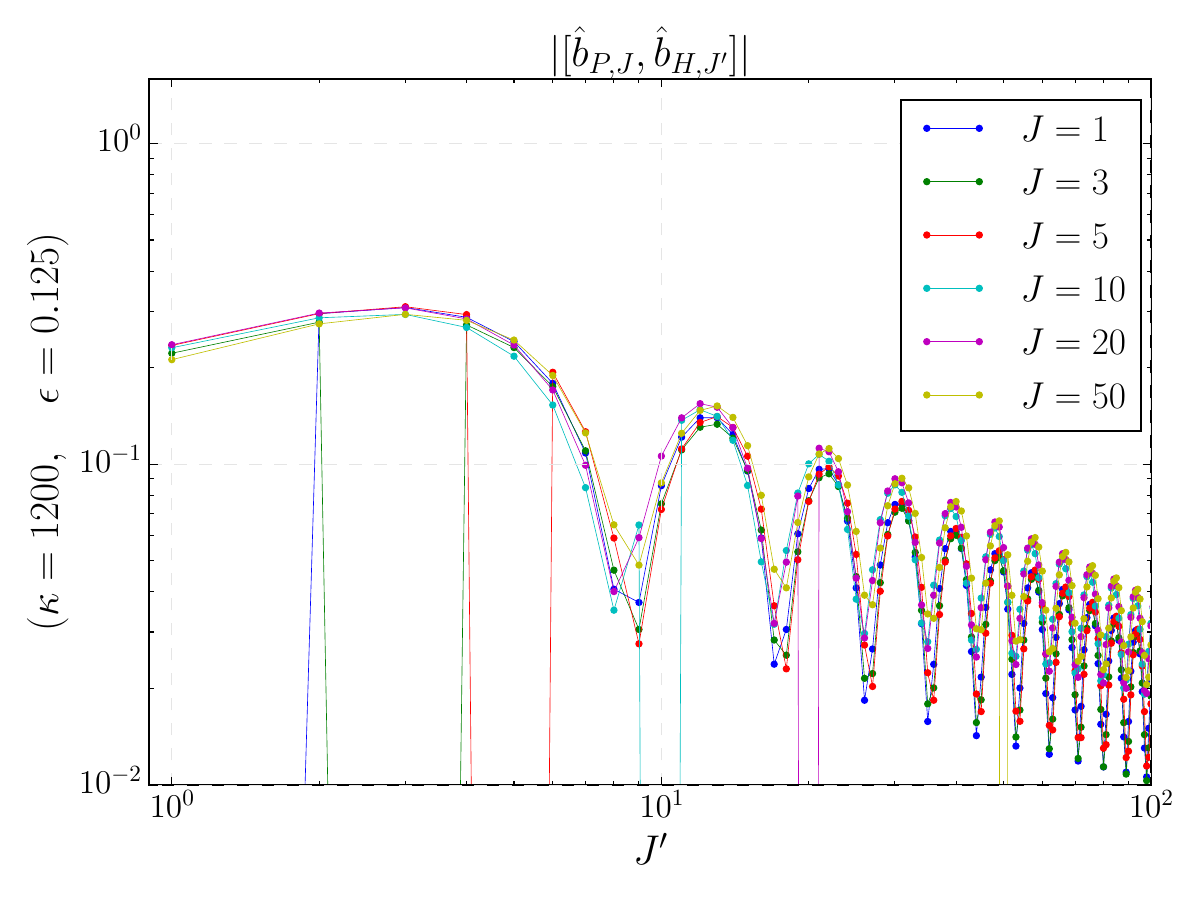}
  \includegraphics[width = 0.48\textwidth]{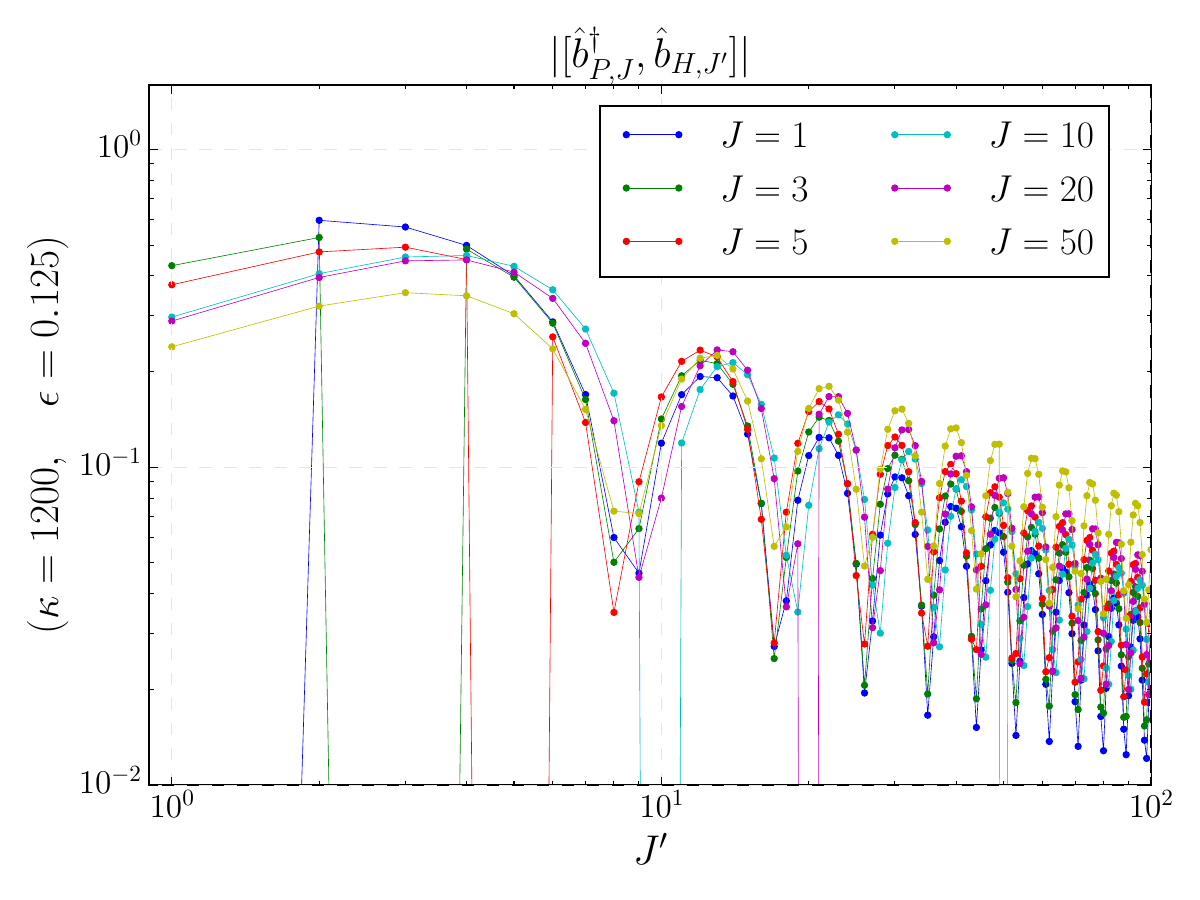}
}
\caption{\justifying Commutators of Hawking modes and partners: these plots correspond to the trajectories given by Eq. \eqref{eq:Traj-1plt}, with $\epsilon=0.125$ and $\kappa=1200$. We show the (modulus of the) commutator in Eq. \eqref{eq:comm-bp-bh} (upper panel) and the one in Eq. \eqref{eq:comm-bpdagg-bh} (lower panel). We see that they obey the usual commutation relations if $J=J'$ (see the nearly vertical lines joining points with vanishingly small values). Otherwise, they do not vanish and are not negligible. This simulation corresponds to $N=1024$. Other choices of $N$ give qualitatively similar results. }
\label{fig:1plt-sharp-HP-comm}
\end{figure}

For the sake of completeness, in Fig. \ref{fig:1plt-sharp-HP-comm} we show the same commutators, this time, for a sharp trajectory of the boundary, namely, for relatively large accelerations. As we see, the situation is more drastic with regard to the nonvanishing values of the commutators for $J\neq J'$.

\begin{figure}[H]
{\centering     
  \includegraphics[width = 0.48\textwidth]{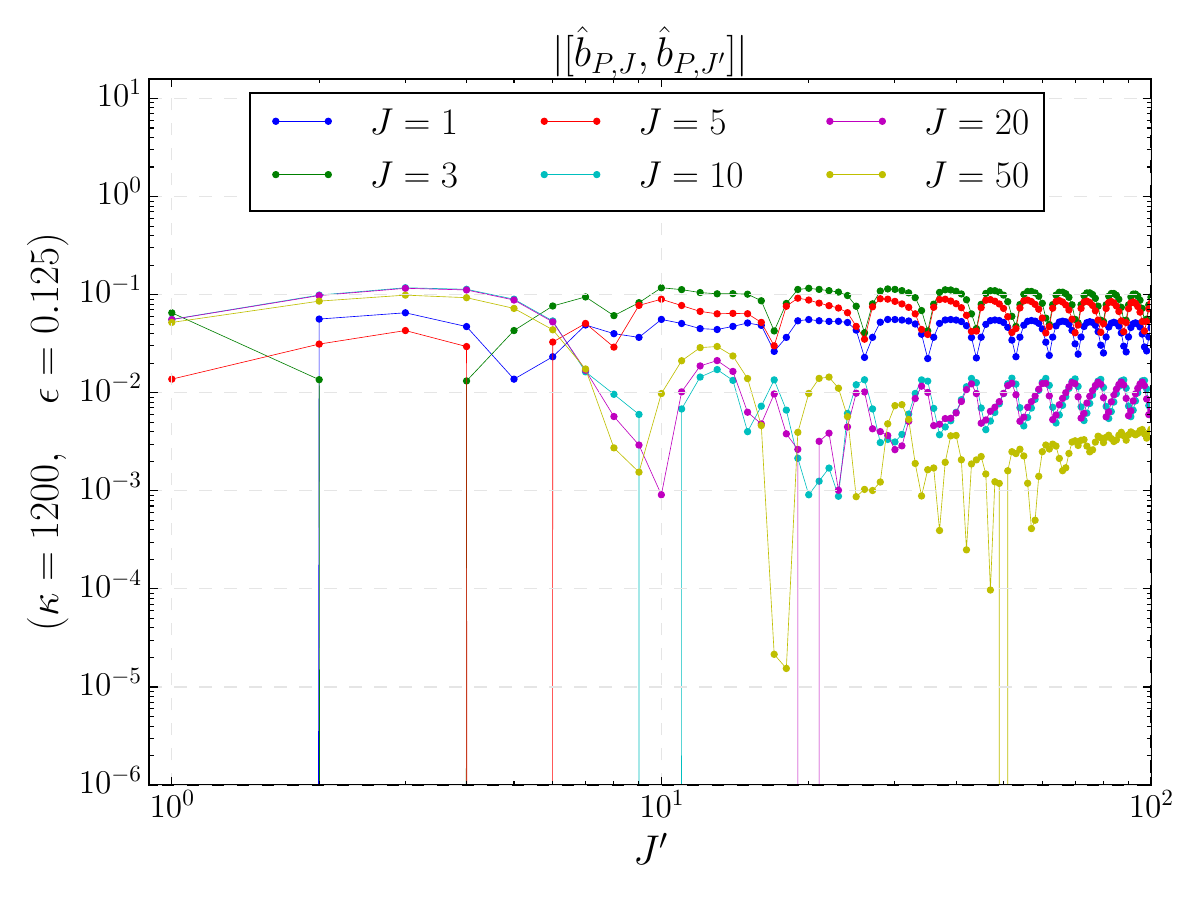}
  \includegraphics[width = 0.48\textwidth]{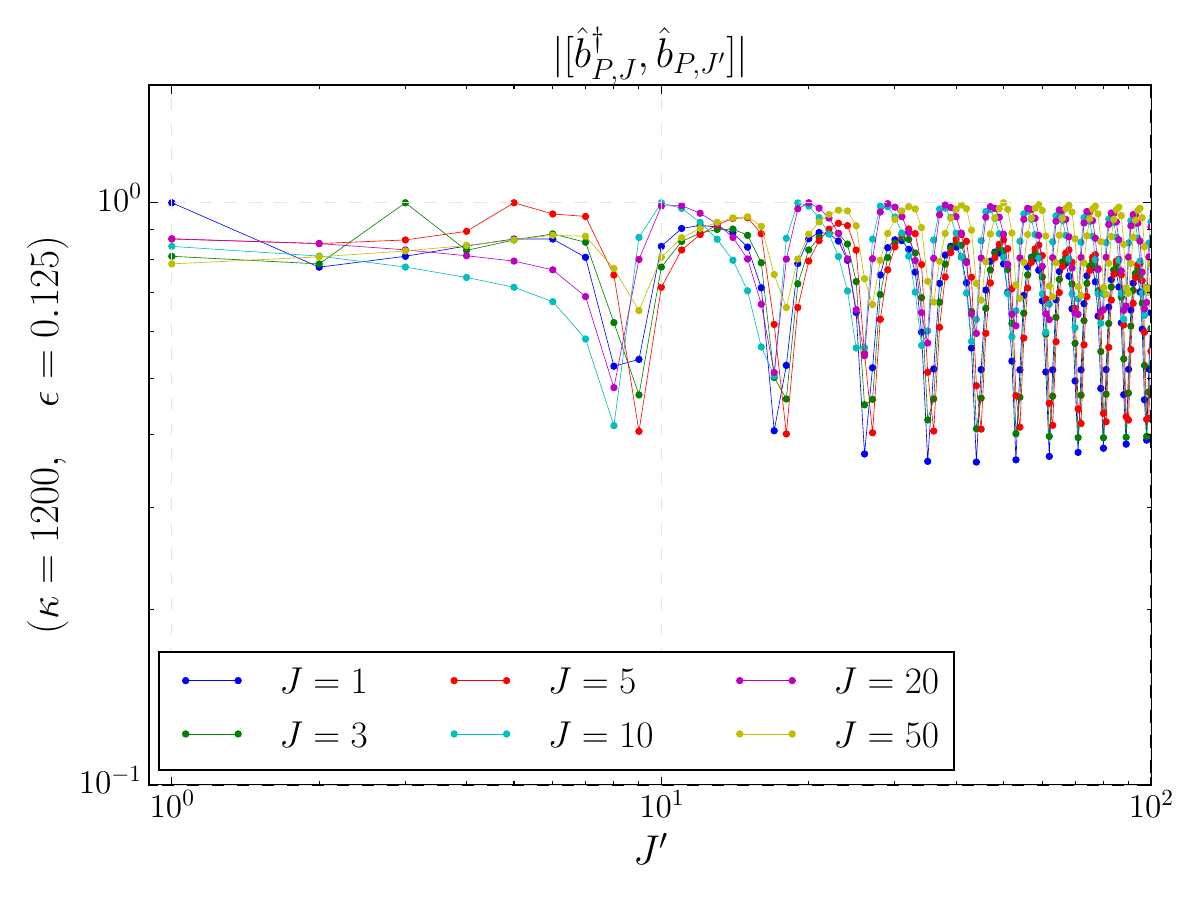}
}
\caption{\justifying Commutators of Hawking modes and partners: these plots correspond to the trajectories given by Eq. \eqref{eq:Traj-1plt}, with $\epsilon=0.125$ and $\kappa=1200$. We show the (modulus of the) commutator in Eq. \eqref{eq:comm-bp-bp} (upper panel) and the one in Eq. \eqref{eq:comm-bpdagg-bp} (lower panel). We see that they obey the usual commutation relations if $J=J'$ (see the nearly vertical lines joining points with vanishingly small values). Otherwise, they do not vanish and are not negligible. This simulation corresponds to $N=1024$. Other choices of $N$ give qualitatively similar results. }
\label{fig:1plt-sharp-PP-comm}
\end{figure}

As in the example above, the commutators in Fig. \ref{fig:1plt-sharp-HP-comm} decrease toward large values of $J'$, while those in Fig. \ref{fig:1plt-sharp-PP-comm} oscillate around constant values of the order of $10^{-1}$ to $10^{0}$.

It would be interesting to apply the concept of quantum information capsule introduced in Refs. \cite{Yamaguchi:2019pnj,Yamaguchi:2020uym} in order to identify  pure-mode subsystems. This will be a matter of future research.

\section{One-by-one ${\rm  LogNeg}$}\label{app:one-by-one-logneg}

\begin{figure}[H]
{\centering     
  \includegraphics[width = 0.48\textwidth]{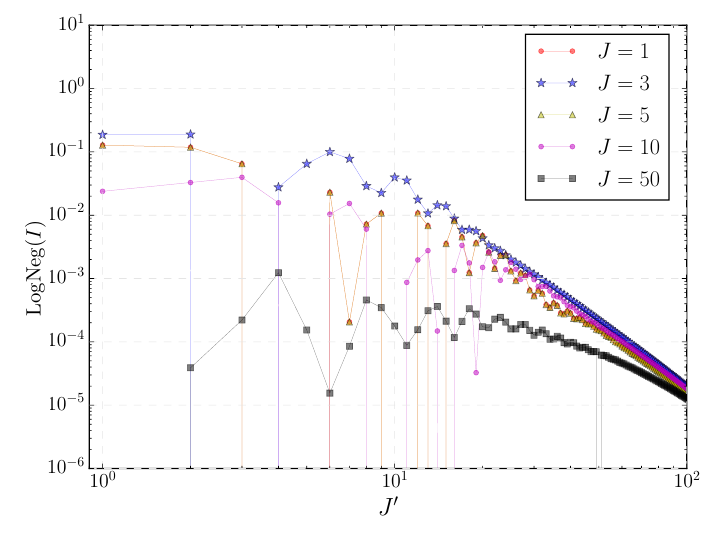}
}
\caption{\justifying One-by-one ${\rm  LogNeg}$: this plot corresponds to the trajectories given by Eq. \eqref{eq:Traj-1plt}, with $\epsilon=0.375$ and $\kappa=33.3$. We show ${\rm  LogNeg}(J,J')$. This simulation corresponds to $N=1024$. }
\label{fig:1plt-one-by-one}
\end{figure}

In this appendix we compute the logarithmic negativity between single Hawking modes. Concretely, we consider bipartitions where subsystem $A$ is a mode $J$ and subsystem $B$ is a given mode $J'$. We then compute the logarithmic negativity between this two subsystems, and denote it by ${\rm  LogNeg}(J,J')$. In Fig. \ref{fig:1plt-one-by-one} we show these quantum correlations for a trajectory of the boundary with relatively small accelerations and for several choices of $J$ and $J'\in[1,100]$.  
We see that quantum entanglement is not uniformly distributed. For instance, for $J=1$, quantum correlations become negligible if $J'$ is large, but remain non negligible for small $J'$, being zero for some bipartitions, for instance $J'=8$ or $J'=12$. However, for $J=3$, quantum correlations are not zero for all bipartitions with $J'\in[1,100]$. On the other hand, if $J$ is large, quantum correlations remain nonvanishing (although they are small) for all values of $I'$ except the smallest ones. For instance, for $J=10$ and $J'=4$ or $J=50$ and $J'=1$, quantum correlations are zero. On the other hand, we see that for $J\geq 3$, ${\rm  LogNeg}(J,J')$ tend to the same asymptote for $J'\gg 1$.

\begin{figure}[H]
{\centering     
  \includegraphics[width = 0.48\textwidth]{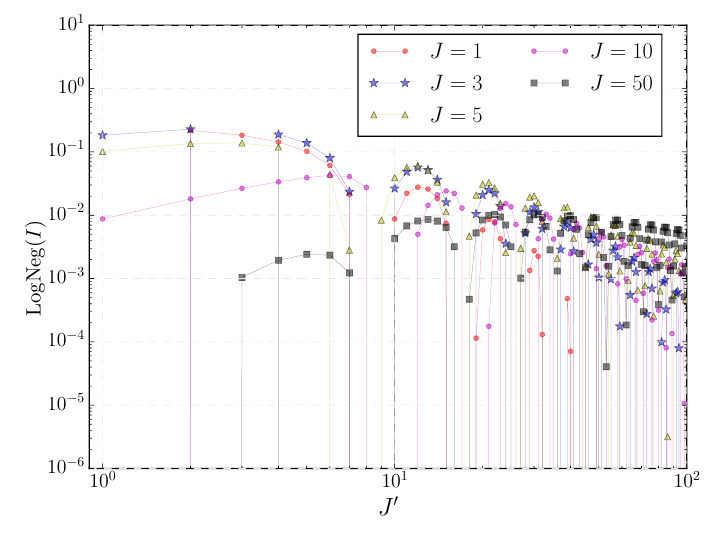}
}
\caption{\justifying One-by-one $\rm LogNeg$: this plot corresponds to the trajectories given by Eq. \eqref{eq:Traj-1plt}, with $\epsilon=0.125$ and $\kappa=1200$. We show ${\rm  LogNeg}(J,J')$. This simulation corresponds to $N=1024$. }
\label{fig:1plt-sharp-one-by-one}
\end{figure}

For the sake of completeness, we also include a simulation for a sharp trajectory of the boundary, i.e., relatively large accelerations, in Fig. \ref{fig:1plt-sharp-one-by-one}. Specifically, we show these quantum correlations for several choices of $J$ and $J'\in[1,100]$. We see that quantum entanglement shows some structure. The quantum entanglement of ${\rm  LogNeg}(J,J')$ is not zero except for some bipartitions, for instance, for $J=1$ and $J'=8,9,16,17,18,\ldots$ or for $J=10$ and $J'=9,11,18,19,20,\ldots\;$. Actually, in all cases, we see that ${\rm  LogNeg}(J,J')$ oscillates with an amplitude that decreases following a power law with $J'$, and such that it always vanishes around the minima of the oscillations. The only exceptions are for the smallest values of $J$, where ${\rm  LogNeg}(J,J')$ vanishes completely once $J'$ reaches a given value. 

\bibliography{referencias}

\end{document}